\documentclass[aps,pra,twocolumn,superscriptaddress,nofootinbib]{revtex4}

\usepackage{graphicx}
\usepackage{dcolumn}
\usepackage{bm}
\usepackage{amssymb}
\usepackage{amsfonts}
\usepackage{latexsym}
\usepackage{enumerate}
\usepackage{color}
\usepackage{amsmath}
\usepackage{epsfig}
\usepackage{times}
\usepackage{algorithm}
\usepackage{algorithmic}
\usepackage{cases}
\usepackage{braket}
\usepackage{framed,color}
\usepackage{CJKutf8}
\usepackage{tikz} %図を描く
\usetikzlibrary{positioning, intersections, calc, arrows.meta,math} %tikzのlibrary

\newtheorem{theorem}{Theorem}

\begin{document}
\widetext
\title{Anonymous quantum sensing robust against state preparation errors}
\author{Hiroto Kasai}
\email{kasai.hiroto.tkb_gw@u.tsukuba.ac.jp}
\affiliation{Graduate School of Pure and Applied Sciences, University of Tsukuba, 1-1-1 Tennodai, Tsukuba, Ibaraki 305-8571, Japan}
\affiliation{Graduate School of Science and Technology, Keio University, Yokohama, Kanagawa 223-8522, Japan }
\author{Seiichiro Tani}
\email{seiichiro.tani@acm.org}
\affiliation{Department of Mathematics, Waseda University, 1-6-1 Nishi-Waseda, Shinjuku-ku, Tokyo 169-8050, Japan}
\author{Yasuhiro Tokura}
\email{tokura.yasuhiro.ft@u.tsukuba.ac.jp}
\affiliation{Graduate School of Pure and Applied Sciences, University of Tsukuba, 1-1-1 Tennodai, Tsukuba, Ibaraki 305-8571, Japan}
\author{Yuki Takeuchi}
\email{Takeuchi.Yuki@bk.MitsubishiElectric.co.jp}
\affiliation{NTT Communication Science Laboratories, NTT Corporation, 3-1 Morinosato Wakamiya, Atsugi, Kanagawa 243-0198, Japan}
\affiliation{NTT Research Center for Theoretical Quantum Information, NTT Corporation, 3-1 Morinosato Wakamiya, Atsugi, Kanagawa 243-0198, Japan}
\affiliation{Information Technology R\&D Center, Mitsubishi Electric Corporation, Kamakura, Kanagawa 247-8501, Japan}
\begin{abstract}
Networked quantum sensors have several applications such as the mapping of magnetic fields.
When the magnetic fields are biomagnetic ones, i.e., they contain some private information, the information of from who non-zero magnetic fields occur has to be protected from eavesdroppers.
Anonymous quantum sensing keeps it secret by estimating amplitudes of the magnetic fields without disclosing the positions of non-zero magnetic fields.
In this paper, we propose an anonymous quantum sensing protocol that is robust against any independent noise in state preparations.
To this end, we devise a quantum state verification protocol for a superposition of Greenberger-Horne-Zeilinger and Dicke states and combine it with the original protocol of anonymous quantum sensing.
Our verification protocol can decide whether the fidelity between the ideal and actual states is high or low more efficiently than the direct fidelity estimation.
Since the original protocol of anonymous quantum sensing cannot correctly estimate the amplitudes of the magnetic fields under state preparation errors, our results would improve the performance of anonymous quantum sensing in realistic situations.
\end{abstract}
\maketitle
\section{Introduction}
A purpose of quantum information processing is to provide the superior performance than classical counterparts.
Quantum sensing is a promising application of quantum information processing and can be applied to measure several physical quantities such as magnetic~\cite{MSHHTCJGTZYWL08,BCKATSKWHKHLBJW08} and electric fields~\cite{DFDNRBWRHJW11} and temperature~\cite{NJDBRWHWBSSSIW13}.
Especially, entanglement-based quantum sensors can estimate amplitudes of magnetic fields more accurately than classical sensors~\cite{HMPEPC97}.
To demonstrate this quantum advantage, several proof-of-principle experiments were already conducted~\cite{LBSBCIJLW04,HBBWP07,NKDBSM11}.
\par
Recently, the application range of quantum sensing is further broadened by locating quantum sensors on nodes in quantum network~\cite{PKD17}.
Such networked quantum sensors can be used for e.g. the mapping of magnetic fields~\cite{SDNANBJW10,BD16} and construction of stable and accurate world clock~\cite{KKBJSYL14}.
Contrary to the versatility, the networked quantum sensors have a security defect when the target physical quantity is a biomagnetic field, i.e., the magnetic field to be sensed includes some private information.
If we measure the biomagnetic field in a standard way, the information of from who it occurs would also be disclosed, and thus the private information is leaked to public.
\par
Anonymous quantum sensing~\cite{KTHMT22} resolves the security defect in networked quantum sensors.
By using symmetric quantum states and measurements, it estimates amplitudes of magnetic fields without disclosing the information of from who non-zero magnetic fields occur.
More precisely, it is cooperatively conducted between three parties: a distributor, $2n$ participants, and an observer.
The purpose is to estimate the amplitudes of non-zero magnetic fields occurred from the $2n$ participants without revealing from who they are occurred.
To this end, the distributor first generates a $2n$-qubit symmetric state, which is a superposition of Greenburger-Horne-Zeilinger (GHZ)~\cite{GHZ89} and Dicke states~\cite{D54}, and sends each qubit to each participant.
Then each of the $2n$ participants interacts the single qubit sent from the distributor with one's own magnetic field and sends the qubit to the observer.
Finally, the observer measures all of the received $2n$ qubits with a symmetric measurement.
By repeating these procedures, the amplitudes of non-zero magnetic fields can be obtained from the measurement outcomes.
On the other hand, since the output probability distribution does not depend on the positions of non-zero magnetic fields due to the symmetry in the protocol, the anonymity is satisfied (for details, see Sec.~\ref{II}).
In the original protocol~\cite{KTHMT22} of anonymous quantum sensing, it was assumed that the state preparation by the distributor is noiseless.
This assumption is crucial in the sense that state preparation errors cause the failure of the estimation of non-zero magnetic fields.
In several quantum cryptographic protocols such as quantum key distribution~\cite{TCKLA14,XWSKSTQML15,PKMCT20} and blind quantum computation~\cite{MF13}, protocols robust against state preparation errors have been proposed.
However, it was open whether such error robustness can be added to anonymous quantum sensing. 
\par
In this paper, we resolve the open problem affirmatively by proposing an anonymous quantum sensing protocol that is robust against any independent noise in the state preparations.
Here, independent means that the noise on each initial state is independently determined.
To this end, we devise a quantum state verification (QSV) protocol for the superposition of $2n$-qubit GHZ and Dicke states and combine it with the original protocol of anonymous quantum sensing.
Our QSV protocol can decide whether the fidelity between the ideal and actual noisy states is high or low more efficiently than the direct fidelity estimation~\cite{FL11}.
For a lower bound $\epsilon(\le 1)$ on the noise strength, our QSV protocol requires $O(\epsilon^{-1})$ samples, while the direct fidelity estimation requires $O(\epsilon^{-2})$ samples (for details, see Secs.~\ref{IIIA} and \ref{IV}).
To devise our QSV protocol, we observe that the superposition of $2n$-qubit GHZ and Dicke states collapses to an $n$-qubit GHZ-like~\cite{LHZ20} or a Dicke state by randomly selecting $n$ qubits and then measuring them in the Pauli-$Z$ basis.
Based on this observation, we propose how to adaptively switch between two existing QSV protocols for GHZ-like~\cite{LHZ20} and Dicke states~\cite{LYSZZ19}.
We also discuss a relation between the sensitivity (i.e., how well the amplitudes of non-zero magnetic fields can be estimated) achieved in our anonymous quantum sensing protocol and the sample complexity (i.e., the required number of samples) of our QSV protocol by varying a free parameter in our anonymous quantum sensing protocol.
\par
The rest of this paper is organized as follows.
In Sec.~\ref{II}, as a preliminary, we review anonymous quantum sensing.
In Sec.~\ref{III}, as another preliminary, we explain the general framework of QSV.
We also introduce concrete QSV protocols for GHZ-like and Dicke states, which will be used as subroutines in our QSV protocol.
In Sec.~\ref{IV}, we devise our QSV protocol for a superposition of GHZ and Dicke states.
In Sec.~\ref{V}, we combine the original protocol of anonymous quantum sensing given in Sec.~\ref{II} and our QSV protocol in Sec.~\ref{IV} to make anonymous quantum sensing robust against state preparation errors.
We also discuss a relation between the sensitivity achieved in our anonymous quantum sensing protocol and the sample complexity of our QSV protocol.
Section~\ref{VI} is devoted to conclusion and discussion.
In Appendices A, B, C, and D, we give proofs of Eqs.~(\ref{M1M2M3}), (\ref{eigenvalueM1_2}), (\ref{eigenvalueM2}), and (\ref{eigenvalueM3}), respectively.
In Appendix E, we show that Eq.~(\ref{eigenvalueM1_2}) is larger than or equal to Eqs.~(\ref{eigenvalueM2}) and (\ref{eigenvalueM3}).
\section{Anonymous quantum sensing}
\label{II}
\begin{figure}[t]
\includegraphics[width=9cm,clip]{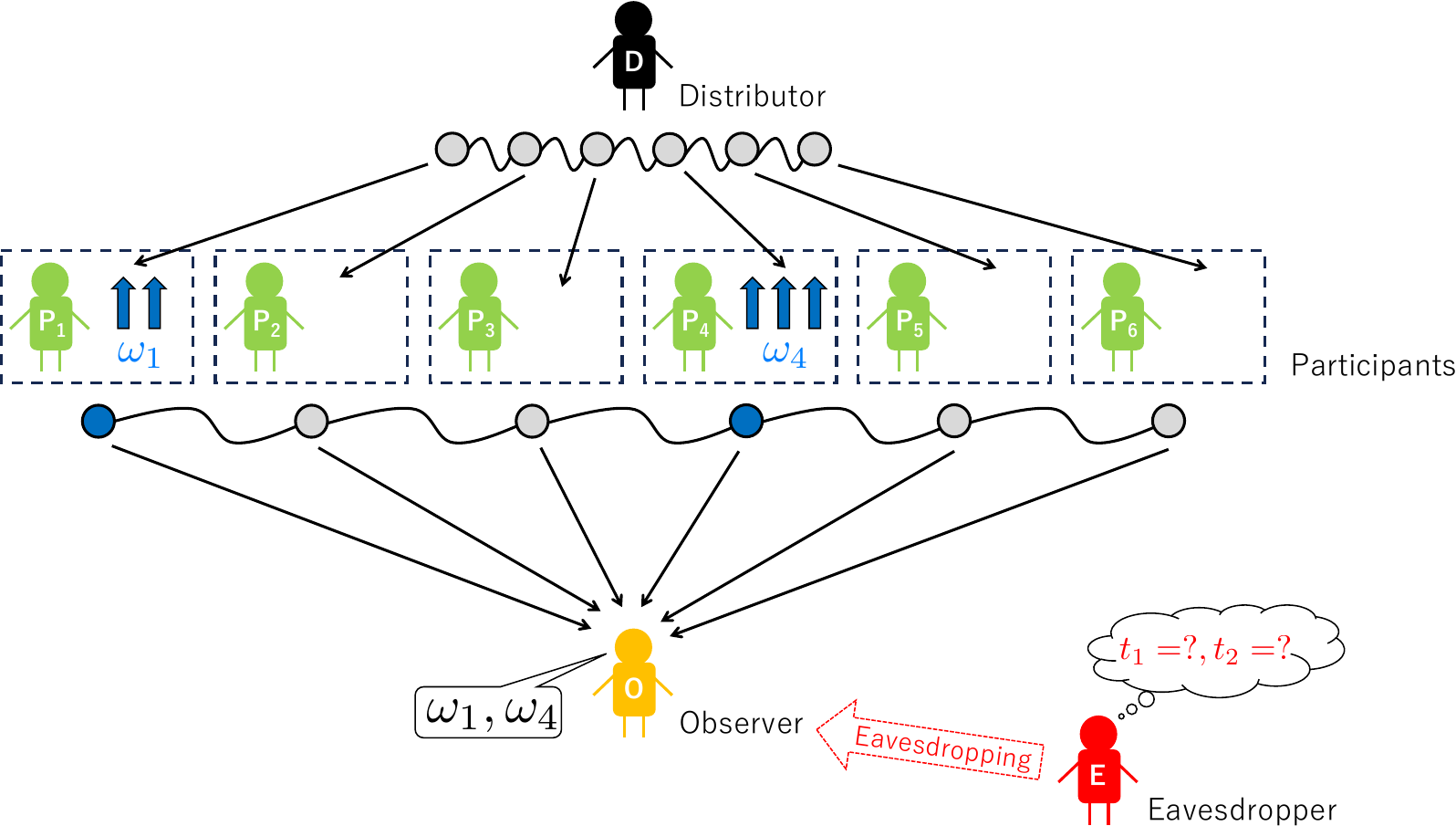}
\caption{Schematic of anonymous quantum sensing.
In this figure, we assume that there are six participants, and non-zero magnetic fields occur from the first and fourth participants, i.e., $t_1=1$ and $t_2=4$.
The distributor prepares the symmetric quantum state in Eq.~(\ref{superposition}) and sends each qubit to each participant.
Then the participants interact their qubits with their own magnetic fields.
As a result, the first and fourth qubits are evolved by the Hamiltonian in Eq.~(\ref{Hamiltonianm}).
Finally, the participants send their qubits to the observer, and the observer measures them with the POVM elements in Eq.~\eqref{eq:povm-of-AQS} to estimate $\omega_1$ and $\omega_4$.
Since the output probability distribution does not depend on $t_1$ and $t_2$, any eavesdropper cannot obtain any information about $t_1$ and $t_2$ even if all classical information is given.}
\label{AQSfig}
\end{figure}

Anonymous quantum sensing~\cite{KTHMT22} is information processing cooperatively conducted between three parties: a distributor, $2n(\ge6)$ participants, and an observer (see Fig.~\ref{AQSfig}).
Note that the distributor and observer can be merged into a single party, but we distinguish them in this paper for clarity.
Its purpose is to estimate amplitudes of non-zero magnetic fields occurred from the $2n$ participants without revealing from who they are occurred.
In other words, by using an anonymous quantum sensing protocol, the observer (e.g., a hospital) can anonymously obtain the amplitudes of non-zero biomagnetic fields of the participants (e.g., patients) with the aid of the distributor (e.g., a quantum state generator).
\par
The original protocol~\cite{KTHMT22} of anonymous quantum sensing proceeds as follows (see also Fig.~\ref{AQSfig}):
\begin{enumerate}
\item The distributor, $2n$ participants, and observer repeat the following procedures $N$ times:
\begin{enumerate}
\item The distributor prepares the $2n$-qubit initial state
\begin{eqnarray}
\label{superposition}
\sqrt{q_0}|{\rm GHZ}_{2n}\rangle+\sqrt{q_1}|D_{2n}^n\rangle,
\end{eqnarray}
where the coefficients $\sqrt{q_0}$ and $\sqrt{q_1}$ are positive real values satisfying $q_0+q_1=1$,
\begin{eqnarray}
|{\rm GHZ}_{2n}\rangle\equiv\cfrac{|0^{2n}\rangle+|1^{2n}\rangle}{\sqrt{2}}
\end{eqnarray}
is the $2n$-qubit GHZ state,
\begin{eqnarray}
|D_{2n}^n\rangle\equiv\cfrac{1}{\sqrt{\binom{2n}{n}}}\sum_{z\in B_{2n,n}}|z\rangle
\end{eqnarray}
is the $2n$-qubit Dicke state with $n$ excitations, and $B_{m,k}$ is the set of all $m$-bit strings having Hamming weight $k$ for any natural number $m$ and non-negative integer $k(\le m)$.
For any $1\le i\le 2n$, the distributor sends the $i$th qubit to the $i$th participant.
\item Each participant interacts the qubit recieved in step (a) with one's own magnetic field for time $t$. As a result, the $i$th qubit is evolved by the Hamiltonian
\begin{eqnarray}
\label{Hamiltonianm}
H_i\equiv\cfrac{\omega_i}{2}Z_i
\end{eqnarray}
for time $t$, where the $i$th resonant frequency $\omega_i(\ge0)$ is proportional to the amplitude of the $i$th participant's magnetic field, and $Z_i\equiv|0\rangle\langle 0|_i-|1\rangle\langle 1|_i$ is the Pauli-$Z$ operator applied to the $i$th qubit.
Then all the participants send their qubits to the observer.
\item The observer performs the positve operator-valued measure (POVM) measurement $\{E^{(j)}\}_{j=1}^4$ on the received $2n$-qubit state, where
\begin{eqnarray}
\label{eq:povm-of-AQS}
\begin{cases}
E^{(1)}
\equiv
|{\rm GHZ}_{2n}\rangle\langle{\rm GHZ}_{2n}|
\\
E^{(2)}
\equiv
Z_1|{\rm GHZ}_{2n}\rangle\langle{\rm GHZ}_{2n}|Z_1
\\
E^{(3)}
\equiv
|D_{2n}^n\rangle\langle D_{2n}^n|
\\
E^{(4)}
\equiv
I^{\otimes 2n}-\sum_{i=1}^3E^{(i)}.
\end{cases}
\end{eqnarray}
Here, $I\equiv|0\rangle\langle 0|+|1\rangle\langle 1|$ is the two-dimensional identity operator.
\end{enumerate}
\item The observer estimates the unknown parameters $\{\omega_i\}_{i=1}^{2n}$ from all the measurement outcomes.
\end{enumerate}
An advantage of this protocol is that no quantum communication between the participants is necessary.
It is worth mentioning that although $q_0|{\rm GHZ}_{2n}\rangle\langle{\rm GHZ}_{2n}|+q_1|D_{2n}^n\rangle\langle D_{2n}^n|$ is used as the initial state in Ref.~\cite{KTHMT22}, we modify it to the superposition in Eq.~(\ref{superposition}).
This modification does not affect the sensitivity and anonymity of the protocol because the coherence between $|{\rm GHZ}_{2n}\rangle$ and $|D_{2n}^n\rangle$ is not used as we can see from Eq.~\eqref{eq:povm-of-AQS}.
On the other hand, it makes the protocol simpler than the original one in the sense that the distributor prepares the same quantum state in each repetition.
Furthermore, as will be explained in Sec.~\ref{IIIA}, this modification is compatible with QSV.
\par
Let us consider a situation where two participants have non-zero magnetic fields, and the amplitudes of the other participants' magnetic fields are zero.
We define two natural numbers $t_1$ and $t_2$ such that the $t_1$th and $t_2$th participants have the non-zero magnetic fields.
We also assume $\omega_{t_1}<\omega_{t_2}\le\pi/(2t)$, which is natural when we are interested in detecting small amplitudes of magnetic fields.
Since the resonant frequency $\omega_i$ is proportional to the amplitude of the $i$th participant's magnetic field, the purpose of anonymous quantum sensing can be restated as obtaining the values of $\omega_{t_1}$ and $\omega_{t_2}$ without revealing the values of $t_1$ and $t_2$.
We explain that this purpose is indeed accomplished by the above protocol.
For simplicity, we define $\theta^+\equiv(\omega_{t_1}+\omega_{t_2})t$ and $\theta^-\equiv(\omega_{t_1}-\omega_{t_2})t$.
As shown in Ref.~\cite{KTHMT22}, the probability $p_j$ of obtaining a measurement outcome that corresponds to the POVM element $E^{(j)}$ is
\begin{eqnarray}
\label{eq:prob-of-outcome-on-AQS}
\begin{cases}
p_1
=
q_0\cfrac{1+\cos{\left(\theta^+\right)}}{2}
\\
p_2
=
q_0\cfrac{1-\cos{\left(\theta^+\right)}}{2}
\\
p_3
=
q_1
\left[
\cfrac{(n-1)\cos{\left(\cfrac{\theta^+}{2}\right)}+n\cos{\left(\cfrac{\theta^-}{2}\right)}}{2n-1}
\right]
^2
\\
p_4
=
1-\sum_{j=1}^3p_j.
\end{cases}
\end{eqnarray}
From Eq.~\eqref{eq:prob-of-outcome-on-AQS}, 
the output probability distribution depends on the values of $\omega_{t_1}$ and $\omega_{t_2}$ but does not depend on the values of $t_1$ and $t_2$.
Thus the protocol has the anonymity.
More precisely, any eavesdropper can obtain no information about $t_1$ and $t_2$ even if all classical information that the distributor, participants, and observer acquire during the protocol is given.
This property is called traceless~\cite{CW05}.
On the other hand,
\begin{eqnarray}
\begin{cases}
\theta^+
=
\cos^{-1}{\left(\cfrac{p_1-p_2}{q_0}\right)}
\\
\theta^-
=
2\cos^{-1}{\left(f(p_1,p_2,p_3)\right)},
\end{cases}
\end{eqnarray}
where
\begin{eqnarray}
{}
&{}&
f(p_1,p_2,p_3)
\nonumber\\
&\equiv&
\cfrac{2n-1}{n}\sqrt{\cfrac{p_3}{q_1}}-\cfrac{n-1}{n}\cos{\left(\cfrac{\cos^{-1}{\left(\cfrac{p_1-p_2}{q_0}\right)}}{2}\right)},
\end{eqnarray}
and hence the values of $\omega_{t_1}$ and $\omega_{t_2}$ can be obtained from the output probability distribution $\{p_i\}_{i=1}^4$.
Note that although $\theta^+$ can be obtained from only $p_1$, we use the POVM element $E^{(2)}$ to improve (an upper bound on) the sensitivity.
In anonymous quantum sensing, the sensitivity is defined as the inverses of the variances $(\Delta\theta^+)^2$ and $(\Delta\theta^-)^2$.
As shown in Ref.~\cite{KTHMT22}, from the Cram\'{e}r-Rao inequality,
\begin{eqnarray}
\label{sensitivity}
\begin{cases}
N(\Delta\theta^+)^2\ge G_{+} 
\\
N(\Delta\theta^-)^2\ge G_{-} 
,
\end{cases}
\end{eqnarray}
where
\begin{eqnarray}
\label{QFI1}
&&
G_{+}
\equiv
\cfrac{1}{q_0}\\
\nonumber
G_{-}&\equiv&\cfrac{1}{1-q_0}+\cfrac{\left(1-\cfrac{1}{n}\right)^2\sin^2{\left(\cfrac{\theta^+}{2}\right)}}{q_0(1-q_0)\sin^2{\left(\cfrac{\theta^-}{2}\right)}}\\
\label{QFI2}
&&+2\cfrac{\left(1-\cfrac{1}{n}\right)\left(1-\cos{\left(\cfrac{\theta^+}{2}\right)}\cos{\left(\cfrac{\theta^-}{2}\right)}\right)}{(1-q_0)\sin^2{\left(\cfrac{\theta^-}{2}\right)}}.
\end{eqnarray}
\par
Since we assume $0<q_0<1$ and $-\pi/2\le\theta^-<0$, the right-hand sides of Eq.~\eqref{sensitivity}  are finite, and thus the lower bounds on $(\Delta\theta^+)^2$ and $(\Delta\theta^-)^2$ can be arbitrarily decreased by increasing the repetition number $N$.
In Ref.~\cite{KTHMT22}, to numerically evaluate $G_-$ in the range of $0<\theta^+\le\pi$ and $0<|\theta^-|\le\pi$, the coefficient $\sqrt{q_0}$ is empirically set as $\sqrt{0.33}$.
\par
In this protocol, all of the state preparation, interaction with the magnetic fields, and POVM measurements are assumed to be noiseless.
Particularly, state preparation errors are serious problems in the sense that if the initial state in Eq.~(\ref{superposition}) is collapsed to $|{\rm GHZ}_{2n}\rangle$, the unknown parameter $\theta^-$ cannot be obtained.
In fact, when the initial state is $|{\rm GHZ}_{2n}\rangle$, the output probabilities $\{p_j\}_{j=1}^4$ become
\begin{eqnarray}
\begin{cases}
p_1
=
\cfrac{1+\cos{(\theta^+)}}{2}
\\
p_2
=
\cfrac{1-\cos{(\theta^+)}}{2}
\\
p_3
=
p_4=0,
\end{cases}
\end{eqnarray}
which depend on only $\theta^+$.
It was open whether anonymous quantum sensing is possible in the situation that the state preparation by the distributor is affected by noises.
\section{Quantum state verification (QSV)}
\label{III}
In this section, we review some existing results on QSV.
In Sec.~\ref{IIIA}, we explain the situation considered in QSV and give the sample complexity of the general QSV protocol.
In Secs.~\ref{IIIB} and \ref{IIIC}, we introduce concrete QSV protocols for GHZ-like and Dicke states, which are subroutines of our QSV protocol in Sec.~\ref{IV}.
\subsection{General framework}
\label{IIIA}
We review the general framework of QSV introduced in Ref.~\cite{PLM18}.
The purpose of QSV is to decide whether a given quantum state is an $n$-qubit ideal target state $|\psi_t\rangle$ or some quantum state that is far from $|\psi_t\rangle$.
It can be formulated as follows: given a lower bound $0<\epsilon\le1$ on the noise strength and $M$ quantum states $\bigotimes_{i=1}^M\rho_i$, decide whether
\begin{enumerate}
\item[(i)] $\rho_i=|\psi_t\rangle\langle\psi_t|$ for all $1\le i\le M$, or
\item[(ii)] $\langle\psi_t|\rho_i|\psi_t\rangle\le1-\epsilon$ for all $i$.
\end{enumerate} 
This formulation should be natural when the noise is independent between the $M$ quantum states.
\par
An usual constraint in QSV is that we must accept $\bigotimes_{i=1}^M\rho_i$ with unit probability in case (i).
It implies that the target state has to be pure.
Furthermore, due to this constraint, all we need to take care is the failure probability $p_{\rm fail}$ of deciding (i) in case (ii).
\par
The general QSV protocol proceeds as follows:
\begin{enumerate}
\item A verifier repeats the following procedures for all $1\le i\le M$:
\begin{enumerate}
\item For the $i$th quantum state $\rho_i$, the verifier selects $j$ from a prespecified set $A$ with probability $q_j$.
\item The verifier performs a binary-outcome projective measurement $\{P_j,I^{\otimes n}-P_j\}$ on $\rho_i$.
Let $0$ and $1$ be the outcomes corresponding to $P_j$ and $I^{\otimes n}-P_j$, respectively.
The outcome $0$ implies that $\rho_i$ is accepted.
\end{enumerate}
\item If all the measurement outcomes are $0$, the verifier accepts $\bigotimes_{i=1}^M\rho_i$, i.e., decides (i).
Otherwise, the verifier rejects it, i.e., decides (ii).
\end{enumerate}
Thus the upper bound on the probability of obtaining the measurement outcome $0$ from the $i$th quantum state $\rho_i$ in case (ii) is
\begin{eqnarray}
\label{pfaili}
\max_{\substack{\rho_i\\ \langle\psi_t|\rho_i|\psi_t\rangle\le1-\epsilon}}{\rm Tr}\left[\Omega\rho_i\right],
\end{eqnarray}
where the maximization is taken over all $\rho_i$ satisfying $\langle\psi_t|\rho_i|\psi_t\rangle\le1-\epsilon$, and
\begin{eqnarray}
\Omega\equiv\sum_{j\in A}q_jP_j
\end{eqnarray}
is the Hermitian operator called strategy.
Since the verifier accepts only when all the measurement outcomes are $0$, from Eq.~(\ref{pfaili}),
\begin{eqnarray}
p_{\rm fail}&\le& \max_{\substack{\rho\\ \langle\psi_t|\rho|\psi_t\rangle\le1-\epsilon}}\left({\rm Tr}\left[\Omega\rho\right]\right)^M\\
\label{pfailupper}
&=&\left(1-\nu\left(\Omega\right)\epsilon\right)^M,
\end{eqnarray}
where $\nu(\Omega)$ is the spectral gap of $\Omega$.
Here, we have used $\langle\psi_t|\Omega|\psi_t\rangle=1$ and $\Omega\le I^{\otimes n}$ to derive Eq.~(\ref{pfailupper}).
Therefore, to satisfy $p_{\rm fail}\le\delta$ for any real value $0<\delta<1$, it is sufficient to set
\begin{eqnarray}
M&=&\left\lceil\cfrac{\log{\delta}}{\log{\left(1-\nu\left(\Omega\right)\epsilon\right)}}\right\rceil\\
\label{sampleupper}
&\le&\left\lceil\cfrac{\log{\delta^{-1}}}{\nu(\Omega)\epsilon}\right\rceil,
\end{eqnarray}
where $\lceil\cdot\rceil$ is the ceiling function.
\par
From Eq.~(\ref{sampleupper}), 
to minimize the sample complexity $M$, we must optimize the strategy $\Omega$ such that its spectral gap is maximized.
Since we assume $\langle\psi_t|\Omega|\psi_t\rangle=1$, the spectral gap $\nu(\Omega)$ is equal to $1-\beta(\Omega)$, where $\beta(\Omega)$ is the second largest eigenvalue of $\Omega$.
It implies that the maximization of $\nu(\Omega)$ is equivalent to the minimization of $\beta(\Omega)$.
Although it is trivially possible by setting $\Omega=|\psi_t\rangle\langle\psi_t|$, 
such strategy requires multi-qubit operations when $|\psi_t\rangle$ is an entangled state.
On the other hand, the participants in anonymous quantum sensing cannot perform multi-qubit operations because they do not perform any quantum communication between them.
In short, 
to construct a QSV protocol for the superposition of GHZ and Dicke states that is compatible with anonymous quantum sensing, 
it would be necessary to find a strategy under the constraint that all projectors $\{P_j\}_{j\in A}$ can be implemented with single-qubit operations and classical communication.
\par
In the above framework, 
case (ii) is defined by using the maximal fidelity, and the verifier's measurements are restricted to projective ones.
In Ref.~\cite{ZH19}, they were generalized to the average fidelity and nonprojective measurements, respectively.
\subsection{GHZ-like states}
\label{IIIB}
In this subsection, we introduce a QSV protocol~\cite{LHZ20} for the $n$-qubit GHZ-like state
\begin{eqnarray}
|\psi_t\rangle=\sqrt{\lambda_0}|0^n\rangle+\sqrt{\lambda_1}|1^n\rangle
\end{eqnarray}
with $\lambda_0\ge\lambda_1>0$ and $\lambda_0+\lambda_1=1$.
This protocol is conducted by $n$ verifiers.
\par
For each $n$-qubit state $\rho_j$ of the given $M$ quantum states $\bigotimes_{j=1}^M\rho_j$, the protocol proceeds as follows:\\
\\
\noindent{\bf [QSV protocol 1]}
\begin{enumerate}
\item The first verifier selects $a=0$ or $a=1$ with probability $p(<1)$ or $1-p$, respectively, and sends the value of $a$ to the other $(n-1)$ verifiers.
If $a=0$, the verifiers proceed to step 2.
On the other hand, if $a=1$, they proceed to step 3.
\item For all $1\le i\le n$, the $i$th verifier measures the $i$th qubit of $\rho_j$ in the Pauli-$Z$ basis and sends the measurement outcome $o_i\in\{0,1\}$ to the first verifier.
If all of $\{o_i\}_{i=1}^n$ are the same, then the first verifier accepts $\rho_j$ and halts the protocol.
Otherwise, the first verifier rejects it and halts the protocol.
\item The first verifier selects $1\le k\le n$ uniformly at random and sends the value of $k$ to the other $(n-1)$ verifiers.
\item For any $i\neq k$, the $i$th verifier chooses $r_i\in\{0,1\}$ uniformly at random.
Then the $i$th qubit is measured in the basis $\{S^{r_i}|+\rangle,S^{r_i}|-\rangle\}$ with $|\pm\rangle\equiv(|0\rangle\pm|1\rangle)/\sqrt{2}$ and $S\equiv|0\rangle\langle 0|+i|1\rangle\langle 1|$, and its measurement outcome $o_i$ and the random bit $r_i$ are sent to the $k$th verifier.
\item The $k$th verifier derives $r_k\in\{0,1\}$ and $o_k\in\{0,1\}$ satisfying $\bigoplus_{i=1}^nr_i=\bigoplus_{i=1}^no_i=0$ and performs the measurement
\begin{eqnarray}
\nonumber
&&\left\{FS^{r_k}Z^{o_k+\left(\sum_{i=1}^nr_i\right)/2}|+\rangle\langle+|Z^{o_k+\left(\sum_{i=1}^nr_i\right)/2}{S^\dag}^{r_k}F,\right. \\
\nonumber
&&\left. I-FS^{r_k}Z^{o_k+\left(\sum_{i=1}^nr_i\right)/2}|+\rangle\langle+|Z^{o_k+\left(\sum_{i=1}^nr_i\right)/2}{S^\dag}^{r_k}F\right\}\\
\label{POVMGHZlike}
&&
\end{eqnarray}
on the $k$th qubit, where $F\equiv\sqrt{2}(\sqrt{\lambda_0}|0\rangle\langle 0|+\sqrt{\lambda_1}|1\rangle\langle 1|)$.
The measurement outcome is sent to the first verifier.
\item The first verifier accepts $\rho_j$ if the received measurement outcome is one corresponding to the first element in Eq.~(\ref{POVMGHZlike}).
Otherwise, the first verifier rejects it.
\end{enumerate}
As we can see from the procedures, this protocol requires only single-qubit operations and classical communication.
It is worth mentioning that although $F$ is not a unitary operator, $FS^{r_k}Z^{o_k+\left(\sum_{i=1}^nr_i\right)/2}|+\rangle$ is a pure state for any $\{r_i\}_{i=1}^n\in\{0,1\}^n$ and $o_k\in\{0,1\}$, and hence the first element in Eq.~(\ref{POVMGHZlike}) is a projector.
\par
Let $\vec{\lambda}\equiv(\lambda_0,\lambda_1)$.
The strategy $\Omega_{\rm GHZ-like}(p,\vec{\lambda})$ of the protocol is
\begin{widetext}
\begin{eqnarray}
\nonumber
&&p\sum_{z=0}^1(|z\rangle\langle z|)^{\otimes n}+\cfrac{1-p}{n}\\
\label{strategyGHZ1}
&&\times\sum_{k=1}^n\left[\cfrac{1}{2^{n-1}}\sum_{r_k,o_k\in\{0,1\}}\sum_{\substack{\{r_i\}_{i\neq k}\\\oplus_{i\neq k}r_i=r_k}}\cfrac{\bigotimes_{i\neq k}I_i+(-1)^{o_k}\bigotimes_{i\neq k}S_i^{r_i}X_i{S_i^\dag}^{r_i}}{2}\otimes F_k\cfrac{I_k+(-1)^{o_k+\left(\sum_{i=1}^nr_i\right)/2}S^{r_k}X_k{S^\dag}^{r_k}}{2}F_k\right]\ \ \ \ \ \ \ \\
&=&p\sum_{z=0}^1(|z\rangle\langle z|)^{\otimes n}+\cfrac{1-p}{n}\sum_{k=1}^n\left[\cfrac{1}{2^{n-1}}\sum_{\substack{\{r_i\}_{i=1}^n\\\oplus_{i=1}^nr_i=0}}F_k\cfrac{I^{\otimes n}+(-1)^{\left(\sum_{i=1}^nr_i\right)/2}\bigotimes_{i=1}^nS_i^{r_i}X_i{S^\dag}^{r_i}}{2}F_k\right]\\
\label{strategyGHZf}
&=&p\sum_{z=0}^1(|z\rangle\langle z|)^{\otimes n}+(1-p)\left\{|\psi_t\rangle\langle\psi_t|+\left[\cfrac{1}{n}\sum_{k=1}^n\left(\bigotimes_{i\neq k}I_i\right)\otimes\left(\sum_{z=0}^1\lambda_z|z\rangle\langle z|_k\right)\right]-\sum_{z=0}^1\lambda_z\left(|z\rangle\langle z|\right)^{\otimes n}\right\},
\end{eqnarray}
\end{widetext}
where the third summation in the second term in Eq.~(\ref{strategyGHZ1}) is taken over all $\{r_i\}_{i=1}^n\setminus\{r_k\}$ satisfying $\oplus_{i=1}^nr_i=0$ for a fixed $r_k$, $X\equiv|1\rangle\langle0|+|0\rangle\langle1|$ is the Pauli-$X$ operator, and the subscripts of operators represent to which qubits the operators are applied.
Here, we have used a result in Ref.~\cite{LHZ20} to derive the last equality.
From Eq.~(\ref{strategyGHZf}), we can see $\langle\psi_t|\Omega_{\rm GHZ-like}(p,\vec{\lambda})|\psi_t\rangle=1$ for any $p$.
The probability $p$ is a free parameter used to maximize the spectral gap of $\Omega_{\rm GHZ-like}(p,\vec{\lambda})$.
A concrete value of $p$ will be determined in Sec.~\ref{IV}.
\par
Although we assumed $\lambda_0\ge\lambda_1$ in the above argument, a similar argument holds even if $\lambda_0<\lambda_1$ by applying the Pauli-$X$ operators to all the measurement bases in the QSV protocol 1.
In other words, when $\lambda_0<\lambda_1$, GHZ-like states can be verified by using $X^{\otimes n}\Omega_{\rm GHZ-like}(p,\vec{\lambda})X^{\otimes n}$ as the strategy.
This is because the coefficients of $|0^n\rangle$ and $|1^n\rangle$ can be swapped by applying $X^{\otimes n}$.
\subsection{Dicke states}
\label{IIIC}
In this section, we introduce a QSV protocol~\cite{LYSZZ19} for the $n(\ge3)$-qubit Dicke state with $k$ excitations
\begin{eqnarray}
|\psi_t\rangle=|D_n^k\rangle\equiv\cfrac{1}{\sqrt{\binom{n}{k}}}\sum_{z\in B_{n,k}}|z\rangle,
\end{eqnarray}
where $1\le k\le n-1$.
Note that although the protocol was proposed for $n\ge 4$ and $2\le k\le n-2$ in Ref.~\cite{LYSZZ19}, we can observe that it works even when 
$ 3 \le n $ and 
$k\in\{1,n-1\}$ by referring the analysis of a QSV protocol~\cite{LYSZZ19} for W states~\cite{DVC00}, 
which are Dicke states with a single excitation.
For each $n$-qubit $\rho_j$, the protocol proceeds as follows:\\
\\
\noindent{\bf [QSV protocol 2]}
\begin{enumerate}
\item The first verifier chooses $(k_1,k_2)$ such that $1\le k_1<k_2\le n$ uniformly at random and sends their values to the other $(n-1)$ verifiers.
\item For any $i\notin\{k_1,k_2\}$, the $i$th verifier measures the $i$th qubit of $\rho_j$ in the Pauli-$Z$ basis and sends the measurement outcome $o_i\in\{0,1\}$ to the $k_1$th and $k_2$th verifiers.
\item When $\sum_{i\notin\{k_1,k_2\}}o_i=k$ or $k-2$, the $k_1$th and $k_2$th verifiers measure their qubits in the Pauli-$Z$ basis and send all the measurement outcomes $\{o_i\}_{i=1}^n$ to the first verifier.
If $\sum_{i\notin\{k_1,k_2\}}o_i=k-1$, they measure their qubits in the Pauli-$X$ basis and send all the measurement outcomes to the first verifier.
Otherwise, they just send the measurement outcomes $\{o_i\}_{i\notin\{k_1,k_2\}}$ to the first verifier.
\item In the case of $\sum_{i\notin\{k_1,k_2\}}o_i=k$ or $k-2$, the first verifier accepts $\rho_j$ if $\sum_{i=1}^no_i=k$.
In the case of $\sum_{i\notin\{k_1,k_2\}}o_i=k-1$, the first verifier accepts it only when $o_{k_1}=o_{k_2}$.
Otherwise, the first verifier always rejects it.
\end{enumerate}
As with the QSV protocol for GHZ-like states, this protocol can also be conducted with only single-qubit operations and classical communication.
\par
Let $\tau^+$ and $\tau^-$ be projectors onto the eigenspaces of a tensor product $\tau$ of Pauli operators with eigenvalues $+1$ and $-1$, respectively.
For any set $A$ of qubits, we also define $\mathcal{Z}_A^m\equiv\sum_{z\in B_{|A|,m}}|z\rangle\langle z|_A$, where $|A|$ is the number of qubits in the set $A$.
We omit the subscript $A$ when $A$ includes all qubits and assume that $\mathcal{Z}_A^m=0$ when $m$ is negative.
The strategy $\Omega_{\rm Dicke}(k)$ of the protocol is
\begin{widetext}
\begin{eqnarray}
\label{strategyD1}
&&\cfrac{1}{\binom{n}{2}}\sum_{k_1<k_2}\left[\mathcal{Z}_{\overline{\{k_1,k_2\}}}^k\otimes Z^+_{k_1}\otimes Z^+_{k_2}+\mathcal{Z}_{\overline{\{k_1,k_2\}}}^{k-2}\otimes Z^-_{k_1}\otimes Z^-_{k_2}+\mathcal{Z}_{\overline{\{k_1,k_2\}}}^{k-1}\otimes (X_{k_1}\otimes X_{k_2})^+\right]\\
\nonumber
&=&\cfrac{\left[n(n-1)-k(n-k)\right]\mathcal{Z}^k+\left(\sum_{\substack{u,v\in B_{n,k}\\ u\oplus v\in B_{n,2}}}|u\rangle\langle v|\right)+\binom{n-k+1}{2}\mathcal{Z}^{k-1}+\binom{k+1}{2}\mathcal{Z}^{k+1}+\left[\sum_{\substack{u\in B_{n,k-1}\\ v\in B_{n,k+1}\\ u\oplus v\in B_{n,2}}}\left(|u\rangle\langle v|+|v\rangle\langle u|\right)\right]}{n(n-1)},\\
\label{strategyDf}
&&
\end{eqnarray}
\end{widetext}
where the summation in Eq.~(\ref{strategyD1}) is taken over all $(k_1,k_2)$ such that $1\le k_1<k_2\le n$, $\overline{A}$ represents the complement of a set $A$, and $u\oplus v$ is the bitwise summation of $u$ and $v$.
Here, we have used a result in Ref.~\cite{LYSZZ19} to obtain the equality.
Since the second term
\begin{eqnarray}
J(n,k)\equiv\sum_{\substack{u,v\in B_{n,k}\\ u\oplus v\in B_{n,2}}}|u\rangle\langle v|
\end{eqnarray}
in the numerator of Eq.~(\ref{strategyDf}) is the adjacency matrix of the Johnson graph, and $|D_n^k\rangle$ is its eigenstate with the maximal eigenvalue $k(n-k)$~\cite{BCN89}, the equality $\langle\psi_t|\Omega_{\rm Dicke}(k)|\psi_t\rangle=1$ holds.
\section{Main result 1: verification of superposition of GHZ and Dicke states}
\label{IV}
In this section, we devise our QSV protocol for the superposition of GHZ and Dicke states in Eq.~(\ref{superposition}) by using the QSV protocols 1 and 2 introduced in Sec.~\ref{III}.
To this end, we observe that if we measure the first $n$ qubits in the Pauli-$Z$ basis, then the remaining $n$ qubits become a GHZ-like or Dicke state.
For simplicity, we define
\begin{eqnarray}
\label{lambda02}
\lambda_0&\equiv&\cfrac{\binom{2n}{n}q_0}{\binom{2n}{n}q_0+2q_1}\\
\label{lambda12}
\lambda_1&\equiv&\cfrac{2q_1}{\binom{2n}{n}q_0+2q_1}.
\end{eqnarray}
When the measurement outcome is $0^n$, the remaining state is
\begin{eqnarray}
\sqrt{\lambda_0}|0^n\rangle+\sqrt{\lambda_1}|1^n\rangle
\end{eqnarray}
because $\sqrt{2}(\langle 0^n|\otimes I^{\otimes n})|{\rm GHZ}_{2n}\rangle=|0^n\rangle$ and $\sqrt{\binom{2n}{n}}(\langle 0^n|\otimes I^{\otimes n})|D_{2n}^n\rangle=|1^n\rangle$.
For a similar reason, when the measurement outcome is $1^n$, the remaining state is
\begin{eqnarray}
\sqrt{\lambda_1}|0^n\rangle+\sqrt{\lambda_0}|1^n\rangle=X^{\otimes n}\left(\sqrt{\lambda_0}|0^n\rangle+\sqrt{\lambda_1}|1^n\rangle\right).
\end{eqnarray}
To consider other measurement outcomes, we use the property that
for any subset $R$ of $n$ qubits,
\begin{eqnarray}
|D_{2n}^n\rangle=\cfrac{1}{\sqrt{\binom{2n}{n}}}\sum_{l=0}^n\binom{n}{l}|D_n^l\rangle_R\otimes|D_n^{n-l}\rangle_{\overline{R}}.
\end{eqnarray}
Thus when the Hamming weight of the measurement outcome is $1\le l\le n-1$, the remaining state is $|D_n^{n-l}\rangle$.
The same argument holds even if we first measure other $n$ qubits because the GHZ and Dicke states are invariant under any permutation of qubits.
\par
Here, we assume $q_0\ge2/(\binom{2n}{n}+2)$ to satisfy $\lambda_0\ge\lambda_1$ for Eqs.~(\ref{lambda02}) and (\ref{lambda12}).
From the above observation, we construct our QSV protocol as follows:
\begin{enumerate}
\item $2n$ verifiers repeat the following procedures for all $\{\rho_j\}_{j=1}^M$:
\begin{enumerate}
\item The first verifier selects a subset $R$ of $n$ qubits uniformly at random and sends $R$ to other $(2n-1)$ verifiers.
\item For any $i\in R$, the $i$th verifier measures the $i$th qubit in the Pauli-$Z$ basis and sends the measurement outcome $o_i\in\{0,1\}$ to the verifiers in the set $\overline{R}$.
\item Depending on the value of $\sum_{i\in R}o_i$, the $n$ verifiers in $\overline{R}$ perform one of the following procedures:
\begin{enumerate}
\item When $\sum_{i\in R}o_i=0$, they perform the QSV protocol 1 with $\lambda_0$ in Eq.~(\ref{lambda02}) and $\lambda_1$ in Eq.~(\ref{lambda12}).
\item When $1\le\sum_{i\in R}o_i\le n-1$, they perform the QSV protocol 2 with the excitation $k=n-\sum_{i\in R}o_i$.
\item When $\sum_{i\in R}o_i=n$, they perform the same procedures of the QSV protocol 1 except for that $X$ is additionally applied to all measurement bases. 
In other words, 
if the original $n$-qubit measurement basis is $M$, 
the verifiers use $ X^{\otimes n} M X^{\otimes n} $ as the actual measurement basis.
\end{enumerate}
\item The first verifier receives whether $\rho_j$ is accepted.
\end{enumerate}
\item If all $\rho_j$ is accepted, then the first verifier accepts $\bigotimes_{j=1}^M\rho_j$.
Otherwise, the first verifier rejects it.
\end{enumerate}
\par
From the above procedures, the strategy $\Omega$ of our protocol is
\begin{eqnarray}
\label{strategyfirst}
\Omega=\cfrac{1}{\binom{2n}{n}}\sum_{R}\Omega_R,
\end{eqnarray}
where
\begin{eqnarray}
\nonumber
\Omega_R&\equiv&\mathcal{Z}_R^0\otimes\Omega_{{\rm GHZ-like},\overline{R}}\left(p,\vec{\lambda}\right)\\
\nonumber
&&+\mathcal{Z}_R^n\otimes\Omega_{{\rm GHZ-like},\overline{R}}\left(p,(1,1)-\vec{\lambda}\right)\\
&&+\sum_{l=1}^{n-1}\mathcal{Z}_R^l\otimes\Omega_{{\rm Dicke},\overline{R}}(n-l).
\end{eqnarray}
As shown in Appendix A, the strategy $\Omega$ can be written as a summation of three orthogonal Hermitian operators $\Omega^{(1)}$, $\Omega^{(2)}$, and $\Omega^{(3)}$, i.e.,
\begin{eqnarray}
\label{M1M2M3}
\Omega=\Omega^{(1)}+\Omega^{(2)}+\Omega^{(3)},
\end{eqnarray}
where
\begin{widetext}
\begin{eqnarray}
\nonumber
\Omega^{(1)}&\equiv&\left[p+(1-p)\lambda_0\right]\left(\mathcal{Z}^0+\mathcal{Z}^{2n}\right)+\left[\cfrac{3n-2}{2(2n-1)}-\cfrac{2(1-p)\lambda_0}{\binom{2n}{n}}\right]\mathcal{Z}^n+\cfrac{J(2n,n)}{2n(2n-1)}
\\
&&
+(1-p)
\sqrt{\cfrac{2\lambda_{0}\lambda_{1}}{\binom{2n}{n}}}
\left(\ket{\rm{GHZ}_{2n}}\bra{D_{2n}^n}+\ket{D_{2n}^n}\bra{\rm{GHZ}_{2n}}\right),
\end{eqnarray}
\begin{eqnarray}
\nonumber
&&\Omega^{(2)}\equiv\left(n+1\right)\left\{\cfrac{1}{4(2n-1)}+\cfrac{1-p}{\binom{2n}{n}}\left[1-\cfrac{1}{n}-\left(1-\cfrac{2}{n}\right)\lambda_0\right]\right\}\left(\mathcal{Z}^{n-1}+\mathcal{Z}^{n+1}\right)+\cfrac{1}{2n(2n-1)}\sum_{\substack{u\in B_{2n,n-1}\\ v\in B_{2n,n+1}\\ u\oplus v\in B_{2n,2}}}\left(|u\rangle\langle v|+|v\rangle\langle u|\right),\\
&&
\end{eqnarray}
and
\begin{eqnarray}
\label{eq:Omega3}
\Omega^{(3)}\equiv\cfrac{1-p}{\binom{2n}{n}}\sum_{l=1}^{n-2}\binom{2n-l}{n}\left[\cfrac{l}{n}+\left(1-\cfrac{2l}{n}\right)\lambda_0\right]\left(\mathcal{Z}^l+\mathcal{Z}^{2n-l}\right).
\end{eqnarray}
\end{widetext}
For $|\psi_t\rangle=\sqrt{q_0}|{\rm GHZ}_{2n}\rangle+\sqrt{q_1}|D_{2n}^n\rangle$, we can show $\langle\psi_t|\Omega|\psi_t\rangle=1$.
Since $\Omega^{(2)}|\psi_t\rangle=\Omega^{(3)}|\psi_t\rangle=0$,
\begin{eqnarray}
\nonumber
&&\Omega|\psi_t\rangle\\
&=&\Omega^{(1)}|\psi_t\rangle\\
\nonumber
&=&\left\{\sqrt{q_0}\left[p+(1-p)\lambda_0\right]+\sqrt{q_1}(1-p)\sqrt{\cfrac{2\lambda_0\lambda_1}{\binom{2n}{n}}}\right\}|{\rm GHZ}_{2n}\rangle\\
\nonumber
&&+\left\{\sqrt{q_0}(1-p)\sqrt{\cfrac{2\lambda_0\lambda_1}{\binom{2n}{n}}}+\sqrt{q_1}\left[1-\cfrac{2(1-p)\lambda_0}{\binom{2n}{n}}\right]\right\}|D_{2n}^n\rangle\\
&&\\
&=&\sqrt{q_0}|{\rm GHZ}_{2n}\rangle+\sqrt{q_1}|D_{\rm 2n}^n\rangle,
\end{eqnarray}
where we have used Eqs.~(\ref{lambda02}) and (\ref{lambda12}) to obtain the last equality, and hence $|\psi_t\rangle$ is accepted by our QSV protocol with unit probability.
\par
To evaluate the sample complexity of our QSV protocol, we derive a lower bound on the spectral gap $\nu(\Omega)$.
To this end, we derive the second largest eigenvalue $\beta(\Omega)$.
Since $\Omega^{(1)}$, $\Omega^{(2)}$, and $\Omega^{(3)}$ are orthogonal with each other, and the target state $|\psi_t\rangle$ is an eigenstate of $\Omega^{(1)}$ with the maximum eigenvalue, $\beta(\Omega)$ is the second largest eigenvalue $\beta(\Omega^{(1)})$ of $\Omega^{(1)}$, the largest eigenvalue of $\Omega^{(2)}$, or that of $\Omega^{(3)}$.
We obtain
\begin{eqnarray}
\nonumber
&&\beta(\Omega^{(1)})\\
\nonumber
&=&\left\{
\begin{array}{ll}
p+(1-p)\lambda_0 
\quad 
\left(
1 
\le
\lambda_{0} 
\bigg(
1 + \frac{2}{ \binom{2n}{n} }
\bigg)
+ 
\frac{1}{(2n-1)(1-p)}
\right)
\\
1-\cfrac{1}{2n-1}
-
\cfrac{2(1-p)\lambda_0}{\binom{2n}{n}} 
\quad 
({\rm otherwise})
\end{array}\right.\\
\label{eigenvalueM1_2}
&&
\end{eqnarray}
in Appendix B.
On the other hand, as shown in Appendices C and D, the largest eigenvalues of $\Omega^{(2)}$ and $\Omega^{(3)}$ are
\begin{eqnarray}
\label{eigenvalueM2}
\cfrac{n+1}{2(2n-1)}+\cfrac{(1-p)(n+1)}{\binom{2n}{n}}\left[1-\cfrac{1}{n}-\left(1-\cfrac{2}{n}\right)\lambda_0\right]
\end{eqnarray}
and
\begin{eqnarray}
\label{eigenvalueM3}
\cfrac{(1-p)[(n-2)\lambda_0+1]}{2n},
\end{eqnarray}
respectively.
Since $\beta(\Omega)$ is the maximum of the eigenvalues in Eqs.~(\ref{eigenvalueM1_2}), (\ref{eigenvalueM2}), and (\ref{eigenvalueM3}), the remaining task is to compare them.
From Appendix E, when $n\ge 3$, the second largest eigenvalue $\beta(\Omega^{(1)})$ is larger than or equal to Eqs.~(\ref{eigenvalueM2}) and (\ref{eigenvalueM3}).
In conclusion, we obtain $\beta(\Omega)=\beta(\Omega^{(1)})$.
\par
Therefore, when $1>\lambda_0(1+2/\binom{2n}{n})+1/[(2n-1)(1-p)]$, the inverse of the spectral gap is
\begin{eqnarray}
\label{isg1}
\cfrac{1}{\nu(\Omega)}=\cfrac{1}{1-\beta(\Omega^{(1)})}=\cfrac{1}{\cfrac{1}{2n-1}+\cfrac{2(1-p)\lambda_0}{\binom{2n}{n}}}< 2n-1.
\nonumber\\
\end{eqnarray}
On the other hand, when $1\le\lambda_0(1+2/\binom{2n}{n})+1/[(2n-1)(1-p)]$, it is
\begin{eqnarray}
\label{isg2}
\cfrac{1}{\nu(\Omega)}=\cfrac{1}{(1-p)\lambda_1}=\cfrac{\binom{2n}{n}q_0+2q_1}{2(1-p)q_1}&<&\cfrac{\cfrac{4^n}{\sqrt{\pi n}}q_0+2q_1}{2(1-p)q_1}\ \ \ \ \ \ \\
\label{isg2order}
&=&O\left(\cfrac{4^n}{\sqrt{n}}\right),
\end{eqnarray}
where we have used Wallis' inequality  to obtain the first inequality.
We have also assumed that $q_0$ and $p$ are constant in Eq.~(\ref{isg2order}).
From Eqs.~(\ref{isg1}) and (\ref{isg2}), we notice that the spectral gap is maximized when $p=0$.
Hence, from Eq.~(\ref{sampleupper}), the sample complexity $M$ of our QSV protocol is
\begin{eqnarray}
\nonumber
&&{\rm max}\left\{\left\lceil\cfrac{(2n-1)\log{\delta^{-1}}}{\epsilon}\right\rceil,\left\lceil\left(\cfrac{q_04^n}{2q_1\sqrt{\pi n}}+1\right)\cfrac{\log{\delta^{-1}}}{\epsilon}\right\rceil\right\}\\
\label{scour}
&&\\
&=&O\left(\cfrac{4^n\log{\delta^{-1}}}{\sqrt{n}\epsilon}\right).
\end{eqnarray}
\par
Our QSV protocol is more efficient than the direct fidelity estimation~\cite{FL11} in the sense that the $\epsilon$-dependence of the direct fidelity estimation is $O(\epsilon^{-2})$.
This difference comes from that the purpose of QSV is to distinguish whether received states are ideal or noisy, while that of the direct fidelity estimation is to obtain an estimated value of the fidelity.
With respect to the $n$-dependence, it is unknown whether our QSV protocol is superior to the direct fidelity estimation.
This is because the direct fidelity estimation requires $O(4^n)$ samples for general $2n$-qubit states, but it can be improved to $O(1/\alpha^2)$ when the target state $|\psi_t\rangle$ satisfies that for any tensor product $\tau$ of Pauli operators, $\langle\psi_t|\tau|\psi_t\rangle=0$ or $|\langle\psi_t|\tau|\psi_t\rangle|\ge\alpha$ holds.
When $\tau=X^{\otimes n}\otimes I^{\otimes n}$, our target state satisfies
\begin{eqnarray}
\label{cvalpha}
\langle\psi_t|\tau|\psi_t\rangle\ge2\sqrt{\cfrac{2q_0q_1}{\binom{2n}{n}}},
\end{eqnarray}
and hence the $n$-dependence of the direct fidelity estimation may be the same as that of our QSV protocol.
If we find $\tau$ whose expected absolute value is smaller than Eq.~(\ref{cvalpha}), then we should be able to say that our QSV protocol is superior to the direct fidelity estimation even with respect to $n$.
\section{Main result 2: anonymous quantum sensing robust against state preparation errors}
\label{V}
In this section, we give our anonymous quantum sensing protocol that is robust against independent state preparation errors.
To this end, we combine the original protocol of anonymous quantum sensing in Sec.~\ref{II} and our QSV protocol in Sec.~\ref{IV}.
In Sec.~\ref{VA}, we explain procedures of our anonymous quantum sensing protocol.
In Sec.~\ref{VB}, we discuss a relation between the sensitivity of our anonymous quantum sensing protocol and the sample complexity of our QSV protocol.
Based on this relation, we numerically find optimal values of $q_0$ for several fixed $\omega_{t_1}$ and $\omega_{t_2}$.
\subsection{Protocol}
\label{VA}
Our anonymous quantum sensing protocol proceeds as follows:
\begin{enumerate}
\item The distributor, $2n(\ge6)$ participants, and observer repeat the following procedures until step (c) is conducted $N$ times:
\begin{enumerate}
\item The distributor prepares the $2n(M+1)$-qubit initial state
\begin{eqnarray}
\bigotimes_{j=1}^{M+1}\rho_j,
\end{eqnarray}
where
\begin{eqnarray}
\rho_j\equiv\mathcal{E}_j\left(\sqrt{q_0}|{\rm GHZ}_{2n}\rangle+\sqrt{q_1}|D_{2n}^n\rangle\right),
\end{eqnarray}
the coefficients $\sqrt{q_0}$ and $\sqrt{q_1}$ are positive real values satisfying $q_0+q_1=1$ and $q_0\ge2/(\binom{2n}{n}+2)$, and the independent noise $\mathcal{E}_j$ is the $2n$-qubit identity map or a completely positive trace-preserving (CPTP) map satisfying $(\sqrt{q_0}\langle{\rm GHZ}_{2n}|+\sqrt{q_1}\langle D_{2n}^n|)\rho_j(\sqrt{q_0}|{\rm GHZ}_{2n}\rangle+\sqrt{q_1}|D_{2n}^n\rangle)\le1-\epsilon$.
The value of $M$ is given in Eq.~(\ref{scour}).
For all $1\le i\le 2n$ and $1\le j\le M+1$, the distributor sends the $i$th qubit of $\rho_j$ to the $i$th participant.
\item By using $M$ states $\{\rho_j\}_{j=1}^M$, the participants perform our QSV protocol to check whether $\mathcal{E}_j(\cdot)=I^{\otimes 2n}(\cdot)I^{\otimes 2n}$ for all $j$ (i.e., case (i) in Sec.~\ref{IIIA}).
If our QSV protocol accepts $\bigotimes_{j=1}^M\rho_j$, then the participants proceed to the next step.
Otherwise, they start over from step (a).
\item By using the remaining state $\rho_{M+1}$, 
the participants and observer perform the same procedures as steps (b) and (c) of the original protocol in Sec.~\ref{II}.
\end{enumerate}
\item The observer estimates the unknown parameters $\{\omega_i\}_{i=1}^{2n}$ from all the measurement outcomes obtained in step (c).
\end{enumerate}
As with the original protocol, no quantum communication between the participants is required in this protocol.
\par
In case (i) in Sec.~\ref{IIIA}, anonymous quantum sensing is conducted with unit probability because the ideal state is certainly accepted by our QSV protocol.
In this case, since the output probability distribution is completely the same as that of the original protocol, our protocol has the traceless and achieves the sensitivity in Eq.~\eqref{sensitivity}.
On the other hand, in case (ii), the participants can notice the noises with arbitrarily high probability $1-\delta$ before they interact their qubits with their magnetic fields.
Therefore, our protocol is robust against any independent state preparation error.
\subsection{Optimization of initial state}
\label{VB}
To optimize our anonymous quantum sensing protocol, we consider what value of $q_0$ is the best.
Note that $q_1$ is uniquely determined if $q_0$ is fixed.
More concretely, we would like to simultaneously maximize and minimize the sensitivity (i.e., $1/G_+$ and $1/G_-$) and sample complexity $M$, respectively.
Since $M$ is minimized by minimizing the second largest eigenvalue $\beta(\Omega)$, our purpose is to minimize all of $G_+$, $G_-$, and $\beta(\Omega)$.
To simplify the optimization, we instead consider the minimization of their multiplication $H(q_0)\equiv G_+G_-\beta(\Omega)|_{p=0}$.
Remember that $\beta(\Omega)$ is minimized by setting $p=0$ as discussed in Sec.~\ref{IV}.
\par
We first focus on the $q_0$-dependence of $\beta(\Omega)|_{p=0}$.
From Eq.~(\ref{lambda02}) and Eq.~\eqref{eigenvalueM1_2} with $p=0$, we can express $\beta(\Omega)|_{p=0}=\beta(\Omega^{(1)})|_{p=0}$ in terms of $q_0$ rather than $\lambda_0$ as follows:
\begin{eqnarray}
\nonumber
&&
\beta(\Omega)|_{p=0}\\
\nonumber
&=&
\left\{
\begin{array}{ll}
1 - \frac{1}{2n-1}
-
\frac
{ 2 q_{0} }
{ 2 + \left(\binom{2n}{n} -2\right) q_{0} }
\quad 
( q_{\rm{min}}(n) \le q_0 < q_{\beta}(n) )
\\
\frac
{ \binom{2n}{n} q_{0} }
{ 2 +\left(\binom{2n}{n}-2\right) q_{0} }
\quad 
( q_{\beta}(n) \le q_0 < 1 ),
\end{array}\right.\\
\label{eigenvalueM1_when_p=0}
&&
\end{eqnarray}
where
\begin{eqnarray}
\label{eq:min-of-range-of-q0}
q_{\rm{min}}(n)
\equiv
\frac
{ 2 }
{ \binom{2n}{n} + 2 },
\end{eqnarray}
\begin{eqnarray}
\label{eq:q0_about_each_case_of_eigenvalueM1_is_same_when_p=0}
q_{\beta}(n)
\equiv
\frac
{ 4(n-1) }
{ \binom{2n}{n} + 8n -6 }.
\end{eqnarray}
Here, 
Eq.\eqref{eq:min-of-range-of-q0}
is the minimum value of the domain of $q_0$.
Note that $q_{\rm{min}}(n)<q_{\beta}(n)$ holds when $n\ge 3$.
Therefore, the $q_0$-dependence of $\beta(\Omega)|_{p=0}$ is
\begin{eqnarray}
\label{eigenvalueM1_when_p=0_derivatived_by_q0}
\frac{\partial}{\partial q_{0}}\beta(\Omega)|_{p=0}
\left\{
\begin{array}{ll}
<0
\quad 
(q_{\rm min}(n)\le q_0<q_{\beta}(n))\\
>0
\quad 
( q_{\beta}(n)\le q_0<1).
\end{array}\right.
\end{eqnarray}
\par
On the other hand, from Eq.~(\ref{QFI1}), we immediately obtain
\begin{eqnarray}
\label{QFI2_by_q0}
\frac{ \partial G_{+} }{ \partial q_{0} }<0.
\end{eqnarray}
With respect to $G_-$, from
\begin{eqnarray}
\frac
{ \partial G_{-} }
{ \partial q_{0}}
=
\frac
{
\gamma \left(
q_{0} + \frac{\eta}{\gamma}
\right)^{2}
-
\eta
\left(
1 + \frac{\eta}{\gamma}
\right)
}
{
n^{2}q_{0}^{2}(1-q_{0})^{2} 
\sin^2{\left(\cfrac{\theta^-}{2}\right)}
}
\label{QFI1_by_q0}
\end{eqnarray}
with positive parameters
\begin{eqnarray}
\nonumber
\begin{cases}
\gamma
\equiv
n^2
\sin^2{\left(\cfrac{\theta^-}{2}\right)}+
2(n^2-n)\left(1 -
\cos{\left(\cfrac{\theta^+}{2}\right)}\cos{\left(\cfrac{\theta^-}{2}\right)}\right)
\\
\eta
\equiv
(n-1)^{2}
\sin^2{\left(\cfrac{\theta^+}{2}\right)},
\end{cases}
\end{eqnarray}
\begin{eqnarray}
\label{eq:parameters_of_qG}    
\end{eqnarray}
we can show
\begin{eqnarray}
\frac
{ \partial G_{-} }
{ \partial q_{0}}
\begin{cases}
< 0 \quad (q_{\rm min}(n)\le q_{0}<q_{G})
\\
=0 \quad (q_{0}=q_G)
\\
>0 \quad (q_{G}<q_{0}<1),
\end{cases}
\end{eqnarray}
where
\begin{eqnarray}
\label{eq:qG}
q_{G}
\equiv
\sqrt{
\frac{\eta}{\gamma}
\bigg(
1 + \frac{\eta}{\gamma}
\bigg)
}
-
\frac{\eta}{\gamma}.
\end{eqnarray}
It implies that $ G_{-}$ takes its minimum value at $q_{0}=q_{G}$.
Hereafter, we consider only $(\theta^+,\theta^-)$ satisfying $q_{\beta}(n)<q_{G}$ as specific examples, which guarantee $q_{\rm min}(n)<q_G$.
\begin{figure}[t]
\includegraphics[height=6cm,width=8.5cm]
{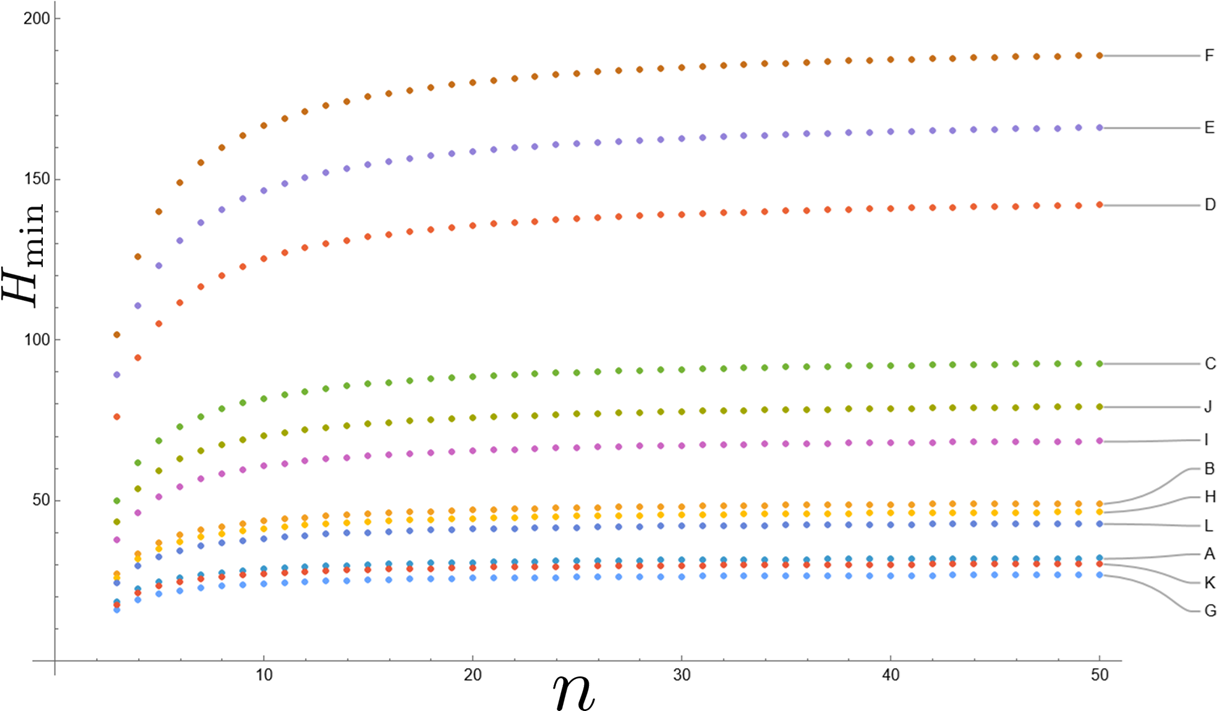}
\caption{
The values of $ H_{\rm{min}} $ for $ 3 \le n \le 50 $.}
\label{fig:Min-of-H-with-n=3-50}
\end{figure}
\begin{figure}[t]
\includegraphics[height=6cm,width=8.5cm]
{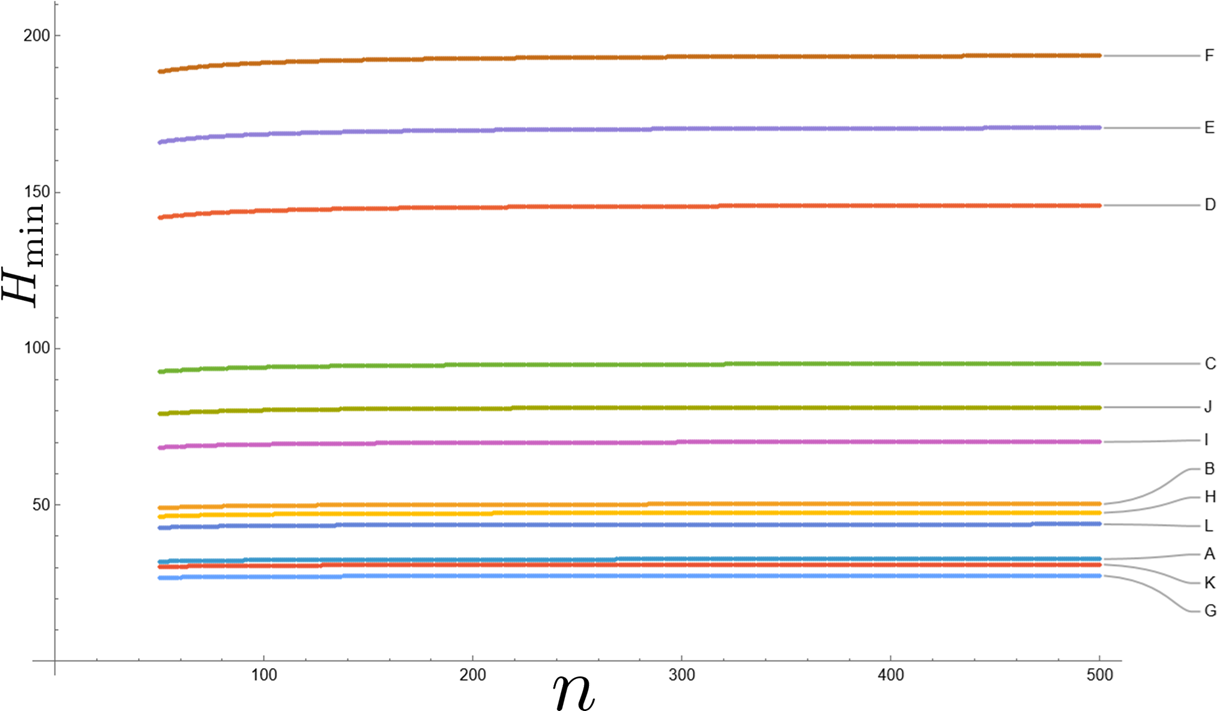}
\caption{
The values of $ H_{\rm{min}} $ for $ 50 \le n \le 500 $.}
\label{fig:Min-of-H-with-n=50-500}
\end{figure}
\par
From Eqs.~(\ref{eigenvalueM1_when_p=0_derivatived_by_q0}) and (\ref{QFI2_by_q0}), we immediately observe that when $q_{\rm min}(n)\le q_0<q_\beta(n)$,
\begin{eqnarray}
\frac{ \partial }{ \partial q_{0} }
( G_{+}\beta(\Omega)|_{p=0} )
<
0.
\end{eqnarray}
The same inequality also holds for $q_\beta(n)\le q_0<1$ because $G_{+}\beta(\Omega)|_{p=0}=\binom{2n}{n}/[2+(\binom{2n}{n}-2)q_0]$.
By combining them with $G_-\ge G_-|_{q_0=q_G}$, the value of $q_H\equiv{\rm argmin}_{q_0}H(q_0)$ is larger than or equal to $q_G$.
\par
By using this observation, we numerically evaluate $q_H$ and $H_{\rm min}\equiv H(q_H)$ for the following examples of $(\theta^+,\theta^-)$:
\begin{eqnarray}
\nonumber
\begin{cases}
A : \left( \frac{\pi}{4}, -\frac{\pi}{6} \right)
,
B : \left( \frac{\pi}{3}, -\frac{\pi}{6} \right)
,
C : \left( \frac{\pi}{2}, -\frac{\pi}{6} \right)
,
D : \left( \frac{2\pi}{3}, -\frac{\pi}{6} \right)
\\
E : \left( \frac{3\pi}{4}, -\frac{\pi}{6} \right)
,
F : \left( \frac{5\pi}{6}, -\frac{\pi}{6} \right)
,
G : \left( \frac{\pi}{3}, -\frac{\pi}{4} \right)
,
H : \left( \frac{\pi}{2}, -\frac{\pi}{4} \right)
\\
I : \left( \frac{2\pi}{3}, -\frac{\pi}{4} \right)
,
J : \left( \frac{3\pi}{4}, -\frac{\pi}{4} \right)
,
K : \left( \frac{\pi}{2}, -\frac{\pi}{3} \right)
,
L : \left( \frac{2\pi}{3}, -\frac{\pi}{3} \right).
\end{cases}
\end{eqnarray}
The $n$-dependence of $ H_{\text{min}} $ is given in Figs.~\ref{fig:Min-of-H-with-n=3-50} and \ref{fig:Min-of-H-with-n=50-500}.
The $n$-dependence in these figures is consistent with the fact that $H(q_0)$ is a monotonically increasing function in $n$.
We also observe that for a fixed $\theta^-$ ($\theta^+$), the minimum value $H_{\rm{min}}$ becomes larger as $\theta^+$ ($\theta^-$) becomes larger.
\par
\begin{figure}[t]
\includegraphics[height=6cm,width=8.5cm]
{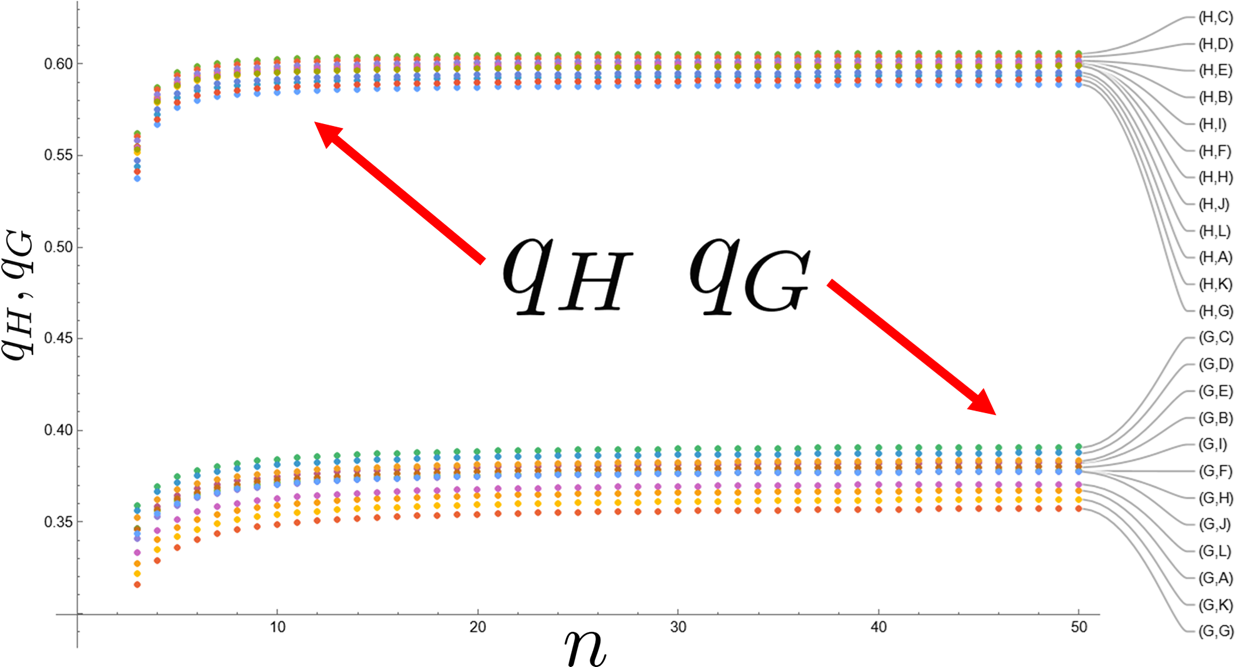}
\caption{The values of $q_H$ and $q_G$ for $3\le n\le50$. For $x\in\{A,B,\cdots,L\}$, the labels $(H,x)$ and $(G,x)$ mean that the corresponding plots represent the values of $q_H$ and $q_G$, respectively.}
\label{figure:qH-and-qG-with-n=3-50}
\end{figure}
\begin{figure}[t]
\includegraphics[height=6cm,width=8.5cm]
{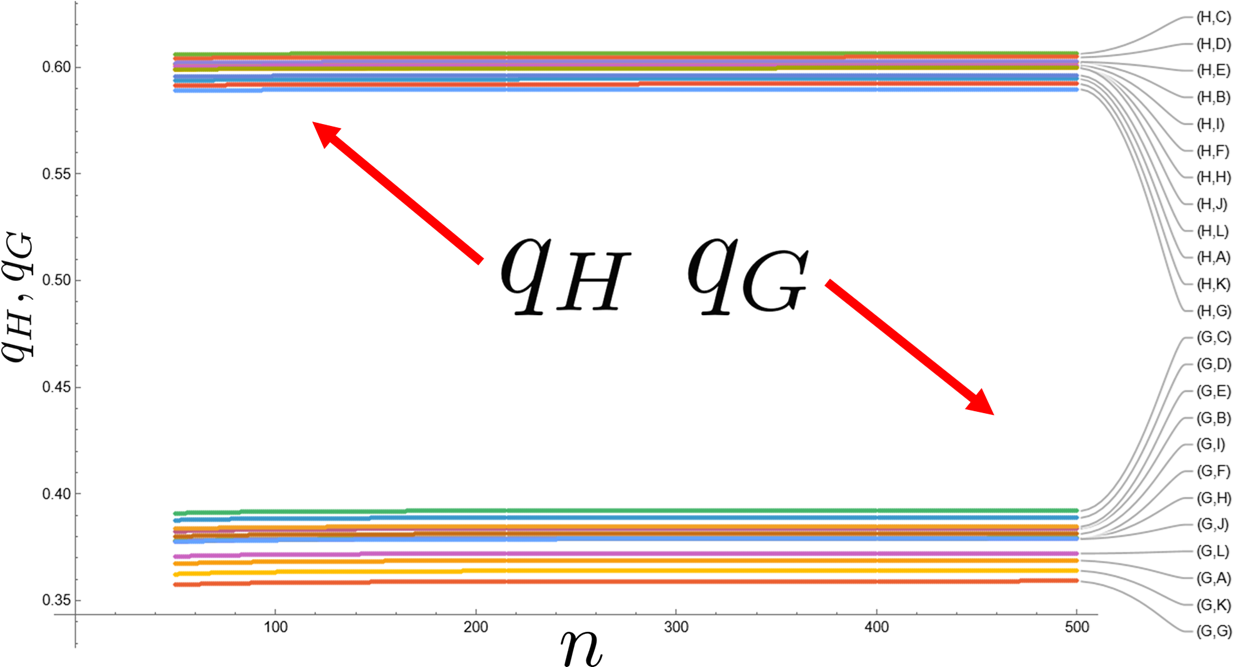}
\caption{The values of $q_H$ and $q_G$ for $50\le n\le500$. For $x\in\{A,B,\cdots,L\}$, the labels $(H,x)$ and $(G,x)$ mean that the corresponding plots represent the values of $q_H$ and $q_G$, respectively.}
\label{figure:qH-and-qG-with-n=50-500}
\end{figure}

The numerical evaluation of $q_H$ is given in Figs.~\ref{figure:qH-and-qG-with-n=3-50} and \ref{figure:qH-and-qG-with-n=50-500} with the values of $q_G$.
Note that the dependence of $q_H$ on $(\theta^+,\theta^-)$ comes from $G_-$.
The discrepancy of $q_H$ and $q_G$ implies that the optimization of a single parameter $G_-$ is not sufficient to optimize the entire performance $H(q_0)$ of anonymous quantum sensing.
The $n$-dependence of $q_G$ in these figures is consistent with that $q_G$ is a monotonically increasing function in $n$.
Similarly, we can observe that $q_H$ is also monotonically increasing with $n$ in our numerical calculation.
As another common property between $q_H$ and $q_G$, these magnitude relations with respect to the $12$ examples from $A$ to $L$ are the same, while they differ from that of $H_{\rm min}$.
\section{Conclusion \& Discussion}
\label{VI}
We have proposed an anonymous quantum sensing protocol that is robust against any independent error in state preparations.
To this end, we have devised a QSV protocol for superpositions of GHZ and Dicke states and have shown that it is more efficient than the direct fidelity estimation~\cite{FL11} in terms of the sample complexity.
We have also derived optimal initial states by evaluating the multiplication of the sensitivity of our anonymous quantum sensing protocol and the sample complexity of our QSV protocol.
Since the optimal initial states depend on the unknown parameters $(\theta^+,\theta^-)$, it seems to be hard to prepare them.
It is open how to prepare a suboptimal initial state.
To further improve the practicality of anonymous quantum sensing, it would be interesting to construct an anonymous quantum sensing protocol that is also robust against both of channel and measurement errors.
\par
Although we have focused on the direct fidelity estimation for comparison, more elaborated verification protocols have also been proposed~\cite{HPS25,SWBF25}.
It would also be interesting to evaluate the performance of their protocols for the superposition of GHZ and Dicke states.
Furthermore, it was recently shown that any pure state can be efficiently verified in an oracle-access model~\cite{GHO25}.
It would be intriguing to combine anonymous quantum sensing with it and optimize the parameter $q_0$.
\section*{ACKNOWLEDGMENTS}
We thank Seiseki Akibue for helpful comments.
We also thank Yuichiro Matsuzaki and Masahiro Takeoka for their comments on the distributor, which make the situation we have in our mind clearer.
H. Kasai is supported by JST Moonshot R\&D-MILLENNIA Program (JPMJMS2061)  and JST Moonshot R\&D (JPMJMS226C).
S. Tani  is supported by JSPS KAKENHI (JP22H00522).
Y. Tokura is partially supported by JST Moonshot R\&D-MILLENNIA Program (JPMJMS2061).
Y. Takeuchi is partially supported by the MEXT Quantum Leap Flagship Program (MEXT Q-LEAP) (JPMXS0120319794) and JST Moonshot R\&D-MILLENNIA Program (JPMJMS2061).

\clearpage
\widetext
\section*{APPENDIX A: PROOF OF EQ.~(\ref{M1M2M3})}
From Eq.~(\ref{strategyfirst}), our strategy can be decomposed as a summation of the GHZ-like component
\begin{eqnarray}
\cfrac{1}{\binom{2n}{n}}\sum_R\left(\mathcal{Z}_R^0\otimes\Omega_{{\rm GHZ-like},\overline{R}}\left(p,\vec{\lambda}\right)+\mathcal{Z}_R^n\otimes\Omega_{{\rm GHZ-like},\overline{R}}\left(p,(1,1)-\vec{\lambda}\right)\right)
\end{eqnarray}
and the Dicke component
\begin{eqnarray}
\cfrac{1}{\binom{2n}{n}}\sum_R\sum_{l=1}^{n-1}\mathcal{Z}_R^l\otimes\Omega_{{\rm Dicke},\overline{R}}(n-l).
\end{eqnarray}
We will calculate each component one by one in the subsections A and B and then will combine them in the subsection C.
\subsection{GHZ-like component}
\label{A1}
We first transform Eq.~\eqref{strategyGHZf} as follows:
\begin{eqnarray}
&{}&
\Omega_{\rm GHZ-like}\left(p,\vec{\lambda}\right)
\nonumber\\
&=&
p\sum_{z=0}^1(|z\rangle\langle z|)^{\otimes n}
+
(1-p)
\left\{
|\psi_t\rangle\langle\psi_t|
+
\left[
\cfrac{1}{n}\sum_{k=1}^n\left(\bigotimes_{i\neq k}I_i\right)
\otimes
\left(\sum_{z=0}^1\lambda_z|z\rangle\langle z|_k\right)
\right]
-
\sum_{z=0}^1\lambda_z\left(|z\rangle\langle z|\right)^{\otimes n}
\right\}\\
&=&
p\sum_{z=0}^1(|z\rangle\langle z|)^{\otimes n}
+
(1-p)
\left[
\lambda_0\mathcal{Z}^0 + \lambda_1\mathcal{Z}^n
+
\sqrt{ \lambda_0 \lambda_1 } 
( \ket{0^n} \bra{1^n} + \ket{1^n} \bra{0^n} )\right]
\nonumber\\
&&+
(1-p)\left\{\lambda_0 \mathcal{Z}^{0} + \lambda_1 \mathcal{Z}^{n}
+
\sum_{a=1}^{n-1} 
\left[
\lambda_0 
+
\cfrac{a( 1 - 2\lambda_0 )}{n} 
\right]
\mathcal{Z}^{a} 
-
\sum_{z=0}^1\lambda_z\left(|z\rangle\langle z|\right)^{\otimes n}
\right\}\\
&=&
p\sum_{z=0}^1(|z\rangle\langle z|)^{\otimes n}
+
(1-p)
\left\{
\sqrt{ \lambda_0 \lambda_1 } 
( \ket{0^n} \bra{1^n} + \ket{1^n} \bra{0^n} )
+
\lambda_0 
\mathcal{Z}^{0}
+
\sum_{a=1}^{n-1} 
\left[
\lambda_0 
+
\cfrac{a( 1 - 2\lambda_0 )}{n} 
\right]
\mathcal{Z}^{a} 
+
\lambda_1
\mathcal{Z}^{n}
\right\}\\
&=&
\label{eq:GHZ-like-strategy}
[ p + (1-p) \lambda_0 ]
\mathcal{Z}^{0}
+
[ p + (1-p) \lambda_1 ] 
\mathcal{Z}^{n}
+
(1-p)
\left\{
\sqrt{ \lambda_0 \lambda_1 } 
( \ket{0^n} \bra{1^n} + \ket{1^n} \bra{0^n} )
+
\sum_{a=1}^{n-1} 
\left[
\lambda_0 
+
\cfrac{a( 1 - 2\lambda_0 ) }{n}
\right]
\mathcal{Z}^{a} 
\right\},
\nonumber\\
\label{strategyGHZliketransform}
\end{eqnarray}
where we have used $|\psi_t\rangle\langle\psi_t|=\lambda_0\mathcal{Z}^0+\lambda_1\mathcal{Z}^n+\sqrt{\lambda_0\lambda_1}(|0^n\rangle\langle 1^n|+\ket{1^n}\bra{0^n})$ and
\begin{eqnarray}
\cfrac{1}{n}\sum_{k=1}^n\left(\bigotimes_{i\neq k}I_i\right)\otimes\left(\sum_{z=0}^1\lambda_z|z\rangle\langle z|_k\right)&=&
\cfrac{1}{n}\sum_{k=1}^n
\left(\bigotimes_{i\neq k}I_i\right)
\otimes
[
\lambda_0 ( I_k - |1\rangle\langle 1|_k )
+
\lambda_1|1\rangle\langle 1|_k
]\\
&=&
\cfrac{1}{n}
\sum_{k=1}^n
\bigg[
\lambda_0I^{\otimes n}
+
( \lambda_1 - \lambda_0 ) 
\left(\bigotimes_{i\neq k}I_i\right)
\otimes
|1\rangle\langle 1|_k 
\bigg]\\
&=&
\lambda_0 I^{\otimes n}
+
\cfrac{\lambda_1 - \lambda_0}{n} 
\sum_{a=1}^{n} 
a \mathcal{Z}^{a}\\
&=&
\lambda_0 
\sum_{a=0}^{n} 
\mathcal{Z}^{a} 
+
\cfrac{1 - 2\lambda_0}{n} 
\sum_{a=1}^{n} 
a \mathcal{Z}^{a}\\
&=&
\lambda_0 \mathcal{Z}^{0} + \lambda_1 \mathcal{Z}^{n}
+
\sum_{a=1}^{n-1} 
\left[
\lambda_0 
+
\cfrac{a( 1 - 2\lambda_0 )}{n} 
\right]
\mathcal{Z}^{a}.
\end{eqnarray}
Note that
\begin{eqnarray}
\sum_{k=1}^{n} 
\left(\bigotimes_{i\neq k}I_i\right)
\otimes
\ket{1}\bra{1}_k
=
\sum_{k=1}^{n} 
\sum_{a=0}^{n-1} 
\mathcal{Z}_{\overline{\{ k \}}}^{a}
\otimes
\ket{1}\bra{1}_k
=
\sum_{a=0}^{n-1} 
(a+1) \mathcal{Z}^{a+1} 
=
\sum_{a=1}^{n} 
a \mathcal{Z}^{a}. 
\end{eqnarray}
Let $S_m$ be a set of all sets of $m$ qubits.
We also define $|A|$ as a number of qubits in the set $A$.
By using Eq.~(\ref{strategyGHZliketransform}), we obtain that for any $m$,
\begin{eqnarray}
&{}&
\sum_{R\in S_m}
\mathcal{Z}_{R}^{0}
\otimes
\Omega_{\rm{GHZ-like},\overline{R}} \left(p,\vec{\lambda}\right)
\nonumber \\
&=&
\sum_{R\in S_m}
\mathcal{Z}_{R}^{0}
\otimes
\left\{
\left[ p + (1-p) \lambda_{0} \right] 
\mathcal{Z}_{\overline{R}}^{0} 
+
[ p + (1-p) \lambda_{1} ]
\mathcal{Z}_{\overline{R}}^{\left|\overline{R}\right|}\right.
\nonumber \\
&&\left. +
(1-p) 
\left\{
\sqrt{ \lambda_{0} \lambda_{1} }
\left( \ket{0^{\left|\overline{R}\right| }} \bra{1^{\left|\overline{R}\right| }} 
+
\ket{1^{\left|\overline{R}\right| }} \bra{0^{\left|\overline{R}\right| }} \right)
+
\sum_{l=1}^{\left|\overline{R}\right|-1} 
\left[
\lambda_{0} 
-
\frac{l( 2 \lambda_{0} -1 )}{\left|\overline{R}\right|}
\right]\mathcal{Z}_{\overline{R}}^{l}
\right\}
\right\}\\
&=&
[ p + (1-p) \lambda_{0} ]
\binom{|V|}{|R|} 
\mathcal{Z}^{0}
+
[ p + (1-p) \lambda_{1} ] 
\mathcal{Z}^{\left|\overline{R}\right|}
\nonumber \\
&&+
(1-p) 
\left\{
\sqrt{ \lambda_{0} \lambda_{1} }
\sum_{v \in B_{|V|,\left|\overline{R}\right|}}
\left( \ket{0^{|V|}} \bra{v} + \ket{v} \bra{0^{|V|}} \right)
+
\sum_{l=1}^{\left|\overline{R}\right|-1} 
\binom{|V|-l}{|R|}
\left[
\lambda_{0} 
-
\frac{l( 2 \lambda_{0} -1 )}{\left|\overline{R}\right|}
\right]\mathcal{Z}^{l}
\right\},
\label{GHZlikedecomposefinal}
\end{eqnarray}
where $V\equiv R\cup\overline{R}$.
To obtain the second equality, we have used that
\begin{eqnarray}
\label{zzcomposition}
\sum_{R\in S_m}\mathcal{Z}_{R}^{a} \otimes \mathcal{Z}_{\overline{R}}^{b}
=
\binom{a+b}{a}
\binom{|V|-(a+b)}{|R|-a} 
\mathcal{Z}^{a+b}
\end{eqnarray}
holds for any $m$, $a(\le m)$, and $b$.
\par
From Eq.~(\ref{GHZlikedecomposefinal}) with $|V|=2n$ and $|R|=m=n$, we finally obtain
\begin{eqnarray}
&&
\nonumber
\sum_{R}
\bigg(
\mathcal{Z}_R^0\otimes\Omega_{{\rm GHZ-like},\overline{R}}\left(p,\vec{\lambda}\right)
+
\mathcal{Z}_R^n\otimes\Omega_{{\rm GHZ-like},\overline{R}}\left(p,(1,1)-\vec{\lambda}\right)
\bigg)
\\
&=&
\sum_{R}
\bigg(
\mathcal{Z}_R^0\otimes\Omega_{{\rm GHZ-like},\overline{R}}\left(p,\vec{\lambda}\right)
+
X^{\otimes 2n}
\mathcal{Z}_R^0\otimes\Omega_{{\rm GHZ-like},\overline{R}}\left(p,\vec{\lambda}\right)
X^{\otimes 2n}
\bigg)
\\
&=&
\sum_{g=0}^1
\left(
X^{\otimes 2n}
\right)
^{g}
\left(
\sum_{R}
\mathcal{Z}_R^0\otimes\Omega_{{\rm GHZ-like},\overline{R}}\left(p,\vec{\lambda}\right)
\right)
\left(
X^{\otimes 2n}
\right)
^{g}
\\
&=&
\sum_{g=0}^1
\left(
X^{\otimes 2n}
\right)
^{g}
\left\{
[ p + (1-p) \lambda_{0} ]
\binom{2n}{n} \mathcal{Z}^{0} 
+
[ p + (1-p) \lambda_{1} ]
\mathcal{Z}^{n}\right. 
\nonumber \\
&&\left.+
(1-p) 
\left\{
\sqrt{ \lambda_{0} \lambda_{1} }
\sum_{ \substack{
v \in B_{2n,n} 
}}
\left( \ket{0^{2n}} \bra{v} + \ket{v} \bra{0^{2n}} \right)
+
\sum_{l=1}^{n-1} 
\binom{2n-l}{n} 
\left[
\lambda_{0} 
-
\frac{l( 2 \lambda_{0} -1 )}{n}
\right]\mathcal{Z}^{l}
\right\}
\right\}
\left(
X^{\otimes 2n}
\right)
^{g}\\
&=&
\label{eq:GHZ_like_component-of-our_state}
[ p + (1-p) \lambda_{0} ]
\binom{2n}{n} 
( \mathcal{Z}^{0} + \mathcal{Z}^{2n} )
+
2 [ p + (1-p) \lambda_{1} ]
\mathcal{Z}^{n} 
\nonumber \\
&&+
(1-p) 
\left\{
\sqrt{ \lambda_{0} \lambda_{1} }
\left(
\sum_{ \substack{
u \in B_{2n,0} \\
v \in B_{2n,n} 
}}
+
\sum_{ \substack{
u \in B_{2n,2n} \\
v \in B_{2n,n} 
}}
\right)
( \ket{u} \bra{v} + \ket{v} \bra{u} )
+
\sum_{l=1}^{n-1} 
\binom{2n-l}{n} 
\left[
\lambda_{0} 
-
\frac{l( 2 \lambda_{0} -1 )}{n}
\right]\left( \mathcal{Z}^{l} + \mathcal{Z}^{2n-l} \right)
\right\}.
\nonumber\\
\end{eqnarray}
\subsection{Dicke component}
\label{A2}
By using
\begin{eqnarray}
\nonumber
\Omega_{{\rm Dicke}}(k)
&=&
\cfrac{
\left[
2\binom{n}{2} - k(n-k) 
\right] 
\mathcal{Z}^{k}
+
\sum_{ \substack{
u,v \in B_{n,k} \\
u \oplus v \in B_{n,2} 
}}
\ket{u} \bra{v}+
\binom{n-k+1}{2} \mathcal{Z}^{k-1}
+
\binom{k+1}{2} \mathcal{Z}^{k+1}
+
\sum_{ \substack{
u \in B_{n,k-1} \\
v \in B_{n,k+1} \\
u \oplus v \in B_{n,2} 
}}
( \ket{u} \bra{v} + \ket{v} \bra{u} )}{n(n-1)},\\
\label{eq:dicke_adaptive_strategy}
\end{eqnarray}
which is give in Eq.~(\ref{strategyDf}) in a slightly different form,
we obtain that for any $m$, $k(\le m)$, and $l(\le k)$,
\begin{eqnarray}
&{}&
\sum_{R\in S_m}
\mathcal{Z}_{R}^{k-l} \otimes 
\Omega_{{\rm Dicke},\overline{R}}(l)
\nonumber \\
&=&
\sum_{R\in S_m}
\mathcal{Z}_{R}^{k-l} \otimes 
\frac{1}{ 2\binom{\left|\overline{R}\right|}{2} }
\left\{
\left[
2\binom{\left|\overline{R}\right|}{2} - l\left(\left|\overline{R}\right|-l\right) 
\right] 
\mathcal{Z}_{\overline{R}}^{l}
+
J\left(\left|\overline{R}\right|,l\right)
+
\binom{\left|\overline{R}\right|-l+1}{2} \mathcal{Z}_{\overline{R}}^{l-1}
+
\binom{l+1}{2} \mathcal{Z}_{\overline{R}}^{l+1}\right.
\nonumber \\
&&\left.+
\sum_{ \substack{
u \in B_{\left|\overline{R}\right|,l-1} \\
v \in B_{\left|\overline{R}\right|,l+1} \\
u \oplus v \in B_{\left|\overline{R}\right|,2} 
}}
( \ket{u} \bra{v} + \ket{v} \bra{u} )
\right\}\\
&=&
\frac{1}{ 2\binom{\left|\overline{R}\right|}{2} }
\left\{
\left[
2\binom{\left|\overline{R}\right|}{2} - l\left(\left|\overline{R}\right|-l\right) 
\right] 
\binom{k}{l}\binom{|V|-k}{\left|\overline{R}\right|-l} 
\mathcal{Z}^{k}
+
\binom{k-1}{l-1}
\binom{|V|-1-k}{\left|\overline{R}\right|-l-1} 
J(|V|,k)\right.
\nonumber \\
&&+
\binom{\left|\overline{R}\right|-l+1}{2}
\binom{k-1}{l-1}
\binom{|V|-(k-1)}{\left|\overline{R}\right|-(l-1)} 
\mathcal{Z}^{k-1}
+
\binom{l+1}{2}
\binom{k+1}{l+1}
\binom{|V|-(k+1)}{\left|\overline{R}\right|-(l+1)} 
\mathcal{Z}^{k+1}
\nonumber \\
&&\left.+
\binom{k-1}{l-1}
\binom{|V|-1-k}{\left|\overline{R}\right|-l-1} 
\sum_{ \substack{
u \in B_{|V|,k-1} \\
v \in B_{|V|,k+1} \\
u \oplus v \in B_{|V|,2} 
}}
( \ket{u} \bra{v} + \ket{v} \bra{u} )
\right\}\\
&=&
\frac{1}{ 2\binom{\left|\overline{R}\right|}{2} }
\left\{
\left[
2\binom{\left|\overline{R}\right|}{2} - l\left(\left|\overline{R}\right|-l\right) 
\right] 
\binom{k}{l}\binom{|V|-k}{\left|\overline{R}\right|-l} 
\mathcal{Z}^{k}\right.
\nonumber \\
&&+
\binom{|V|-k+1}{2}
\binom{|V|-k-1}{\left|\overline{R}\right|-l-1}
\binom{k-1}{l-1} 
\mathcal{Z}^{k-1}
+
\binom{k+1}{2}
\binom{k-1}{l-1}
\binom{|V|-k-1}{\left|\overline{R}\right|-l-1} 
\mathcal{Z}^{k+1}
\nonumber \\
&&\left.+
\binom{k-1}{l-1}
\binom{|V|-1-k}{\left|\overline{R}\right|-l-1} 
\left[
J(|V|,k)+
\sum_{ \substack{
u \in B_{|V|,k-1} \\
v \in B_{|V|,k+1} \\
u \oplus v \in B_{|V|,2} 
}}
( \ket{u} \bra{v} + \ket{v} \bra{u} )
\right]
\right\}\\
&=&
\frac{1}{ 2\binom{\left|\overline{R}\right|}{2} }
\left\{
\left[
2\binom{\left|\overline{R}\right|}{2} - l\left(\left|\overline{R}\right|-l\right) 
\right] 
\binom{k}{l}\binom{|V|-k}{|\overline{R}|-l} 
\mathcal{Z}^{k}\right.
\nonumber \\
&&\left.+
\binom{k-1}{l-1}
\binom{|V|-1-k}{\left|\overline{R}\right|-l-1} 
\left[
\binom{|V|-k+1}{2} 
\mathcal{Z}^{k-1}
+
\binom{k+1}{2} 
\mathcal{Z}^{k+1}
+
J(|V|,k)+
\sum_{ \substack{
u \in B_{|V|,k-1} \\
v \in B_{|V|,k+1} \\
u \oplus v \in B_{|V|,2} 
}}
( \ket{u} \bra{v} + \ket{v} \bra{u} )
\right]
\right\}.
\nonumber \\
\end{eqnarray}
To obtain the second equality, we have used Eq.~(\ref{zzcomposition}) and the two equalities
\begin{eqnarray}
\sum_{R\in S_m}
\mathcal{Z}_{R}^{k-l} \otimes 
J(\left|\overline{R}\right|,l)
&=&
\sum_{R\in S_m}
\mathcal{Z}_{R}^{k-l} \otimes 
\sum_{ \substack{
u,v \in B_{\left|\overline{R}\right|,l} \\
u \oplus v\in B_{\left|\overline{R}\right|,2} 
}}
\ket{u} \bra{v}\\
&=&
\binom{k-1}{l-1}
\binom{|V|-1-k}{\left|\overline{R}\right|-1-l} 
\sum_{ \substack{
u,v \in B_{|V|,k} \\
u \oplus v\in B_{|V|,2} 
}}
\ket{u} \bra{v} 
=
\binom{k-1}{l-1} 
\cdot
\binom{|V|-1-k}{\left|\overline{R}\right|-1-l} 
J(|V|,k)
\end{eqnarray}
and
\begin{eqnarray}
\sum_{R\in S_m}
\mathcal{Z}_{R}^{k-l} \otimes 
\sum_{ \substack{
u \in B_{\left|\overline{R}\right|,l-1} \\
v \in B_{\left|\overline{R}\right|,l+1} \\
u \oplus v\in B_{\left|\overline{R}\right|,2} 
}}
( \ket{u} \bra{v} + \ket{v} \bra{u} )
&=&
\binom{k-1}{l-1}
\binom{|V|-1-k}{\left|\overline{R}\right|-1-l} 
\sum_{ \substack{
u \in B_{|V|,k-1} \\
v \in B_{|V|,k+1} \\
u \oplus v\in B_{|V|,2} 
}}
( \ket{u} \bra{v} + \ket{v} \bra{u} ).
\end{eqnarray}
Therefore, when $|V|=2n$ and $|R|=m=n$,
\begin{eqnarray}
&{}&
\sum_{ R }
\mathcal{Z}_{R}^{n-l} \otimes 
\Omega_{{\rm Dicke},\overline{R}}(l)
\nonumber \\
&=&
\frac{1}{ 2\binom{n}{2} }
\left\{
\left[
2\binom{n}{2} - l(n-l) 
\right] 
\binom{n}{l}\binom{n}{n-l} 
\mathcal{Z}^{n}\right.
\nonumber \\
&&\left.+
\binom{n-1}{l-1}
\binom{n-1}{n-1-l} 
\left[
\binom{n+1}{2} 
\mathcal{Z}^{n-1}
+
\binom{n+1}{2} 
\mathcal{Z}^{n+1}
+
J(2n,n)+
\sum_{ \substack{
u \in B_{2n,n-1} \\
v \in B_{2n,n+1} \\
u \oplus v \in B_{2n,2} 
}}
( \ket{u} \bra{v} + \ket{v} \bra{u} )
\right]
\right\}\\
&=&
\frac{1}{ 2\binom{n}{2} }
\left\{
2\binom{n}{2}\binom{n}{l} 
\left(
\binom{n}{n-l} 
- 
\binom{n-2}{n-1-l} 
\right)
\mathcal{Z}^{n}\right.
\nonumber \\
&&\left.+
\binom{n-1}{n-l}
\binom{n-1}{l} 
\left[
\binom{n+1}{2} 
\left(\mathcal{Z}^{n-1} +\mathcal{Z}^{n+1}\right)
+
J(2n,n)+
\sum_{ \substack{
u \in B_{2n,n-1} \\
v \in B_{2n,n+1} \\
u \oplus v \in B_{2n,2} 
}}
( \ket{u} \bra{v} + \ket{v} \bra{u} )
\right]
\right\}.
\label{Dickedecomposefinal}
\end{eqnarray}
\par
From the Chu–Vandermonde identity~\cite{roman2008umbral}
\begin{eqnarray}
\binom{m+n}{r}
=
\sum_{k=0}^{r}
\binom{m}{k}
\binom{n}{r-k} 
\end{eqnarray}
with any nonnegative integers $m$, $n$, and $r$, we have
\begin{eqnarray}
&{}&
\sum_{l=1}^{ n-1 }
2\binom{n}{2}\binom{n}{l} 
\left(
\binom{n}{n-l} 
- 
\binom{n-2}{n-1-l} 
\right)
\nonumber \\
&=&
2\binom{n}{2} 
\left[
\left(
\sum_{l=0}^{ n }
-
\sum_{l\in\{0,n\}}
\right)
\binom{n}{l}
\binom{n}{n-l} 
- 
\left(
\sum_{l=0}^{ n-1 }
-
\sum_{l=0}
\right)
\binom{n}{l}
\binom{n-2}{n-1-l} 
\right]\\
&=&
2\binom{n}{2} 
\left(
\binom{2n}{n} 
-2
-
\binom{2n-2}{n-1}
\right)\\
&=&
2\binom{n}{2} 
\left[
\binom{2n}{n} 
-
2
- 
\frac{n}{2(2n-1)}
\binom{2n}{n} 
\right]\\
&=&
2\binom{n}{2} 
\left[
\frac{3n-2}{2(2n-1)}
\binom{2n}{n} 
-
2
\right]
\end{eqnarray}
and
\begin{eqnarray}
\sum_{l=1}^{ n-1 }
\binom{n-1}{n-l}
\binom{n-1}{l} 
=
\left(
\sum_{l=0}^{ n }
-
\sum_{l\in\{0,n\}}
\right)
\binom{n-1}{n-l}
\binom{n-1}{l}=
\binom{2n-2}{n}
=
\frac{n-1}{2(2n-1)}
\binom{2n}{n}.
\end{eqnarray}
By using these equalities with Eq.~(\ref{Dickedecomposefinal}), we finally obtain
\begin{eqnarray}
&{}&
\sum_{ R}
\sum_{l=1}^{n-1}
\mathcal{Z}_R^{l}
\otimes
\Omega_{{\rm Dicke},\overline{R}}(n-l)
\nonumber\\
&=&
\sum_{l=1}^{n-1}
\sum_{ R}
\mathcal{Z}_R^{n-l}
\otimes
\Omega_{{\rm Dicke},\overline{R}}(l)\\
&=&
\sum_{l=1}^{ n-1 }
\frac{1}{ 2\binom{n}{2} }
\left\{
2\binom{n}{2}\binom{n}{l} 
\left(
\binom{n}{n-l} 
- 
\binom{n-2}{n-1-l} 
\right)
\mathcal{Z}^{n}\right.
\nonumber \\
&&\left.+
\binom{n-1}{n-l}
\binom{n-1}{l} 
\left[
\binom{n+1}{2} 
\left(\mathcal{Z}^{n-1} + \mathcal{Z}^{n+1}\right)
+
J(2n,n)+
\sum_{ \substack{
u \in B_{2n,n-1} \\
v \in B_{2n,n+1} \\
u \oplus v \in B_{2n,2} 
}}
( \ket{u} \bra{v} + \ket{v} \bra{u} )
\right]
\right\}\\
&=&
\frac{1}{ 2\binom{n}{2} }
\left\{
2\binom{n}{2} 
\left[
\frac{3n-2}{2(2n-1)}
\binom{2n}{n} 
-
2
\right]
\mathcal{Z}^{n}\right.
\nonumber \\
&&\left.+
\frac{n-1}{2(2n-1)}
\binom{2n}{n} 
\left[
\binom{n+1}{2} 
\left(\mathcal{Z}^{n-1} + \mathcal{Z}^{n+1}\right)
+
J(2n,n)+
\sum_{ \substack{
u \in B_{2n,n-1} \\
v \in B_{2n,n+1} \\
u \oplus v \in B_{2n,2} 
}}
( \ket{u} \bra{v} + \ket{v} \bra{u} )
\right]
\right\}\\
&=&
\label{eq:dikce_component-of-our_state}
\left[
\frac{3n-2}{2(2n-1)}
\binom{2n}{n} 
-
2
\right]
\mathcal{Z}^{n}
+
\frac{ \binom{2n}{n} }{2\binom{2n}{2}}
\left[
\binom{n+1}{2} 
\left(\mathcal{Z}^{n-1} + \mathcal{Z}^{n+1}\right)
+
J(2n,n)+
\sum_{ \substack{
u \in B_{2n,n-1} \\
v \in B_{2n,n+1} \\
u \oplus v \in B_{2n,2} 
}}
( \ket{u} \bra{v} + \ket{v} \bra{u} )
\right].\ \ \
\end{eqnarray}
\subsection{Our strategy}
\label{A3}
From
Eqs.
\eqref{eq:GHZ_like_component-of-our_state}
and
\eqref{eq:dikce_component-of-our_state}
,
Eq.~(\ref{strategyfirst}) becomes
\begin{eqnarray}
\Omega
&=&
\frac{1}{ \binom{2n}{n} }
\left\{
[ p + (1-p) \lambda_{0} ]
\binom{2n}{n} 
\left(\mathcal{Z}^{0} + \mathcal{Z}^{2n} \right)
+
2[ p + (1-p) \lambda_{1} ]
\mathcal{Z}^{n}\right.
\nonumber \\
&&+
(1-p) 
\left\{
\sqrt{ \lambda_{0} \lambda_{1} }
\left(
\sum_{ \substack{
u \in B_{2n,0} \\
v \in B_{2n,n} 
}}
+
\sum_{ \substack{
u \in B_{2n,2n} \\
v \in B_{2n,n} 
}}
\right)
( \ket{u} \bra{v} + \ket{v} \bra{u} )
+
\sum_{l=1}^{n-1} 
\binom{2n-l}{n}\left[
\lambda_{0} 
-
\frac{l( 2 \lambda_{0} -1 )}{n}
\right]
\left(\mathcal{Z}^{l} + \mathcal{Z}^{2n-l} \right)
\right\}
\nonumber\\
&&\left.+
\left[
\frac{3n-2}{2(2n-1)}
\binom{2n}{n} 
-
2
\right]
\mathcal{Z}^{n}
+
\frac{ \binom{2n}{n} }{2\binom{2n}{2}}
\left[
\binom{n+1}{2} 
\left(\mathcal{Z}^{n-1} + \mathcal{Z}^{n+1}\right)
+
J(2n,n)+
\sum_{ \substack{
u \in B_{2n,n-1} \\
v \in B_{2n,n+1} \\
u \oplus v \in B_{2n,2} 
}}
( \ket{u} \bra{v} + \ket{v} \bra{u} )
\right]
\right\}\ \ \ \ \ \ \ \ \ \\
&=&
[ p + (1-p) \lambda_{0} ]
\left(\mathcal{Z}^{0} +\mathcal{Z}^{2n}\right)
+
\left[
\frac{3n-2}{2(2n-1)}
-
\frac
{ 2(1-p) \lambda_{0} }
{ \binom{2n}{n} }
\right]
\mathcal{Z}^{n}
+
\frac{ J(2n,n) }{2n(2n-1)}
\nonumber \\
&&+
\frac{
(1-p) 
\sqrt{ \lambda_{0} \lambda_{1} }
}
{ \binom{2n}{n} }
\left(
\sum_{ \substack{
u \in B_{2n,0} \\
v \in B_{2n,n} 
}}
+
\sum_{ \substack{
u \in B_{2n,2n} \\
v \in B_{2n,n} 
}}
\right)
( \ket{u} \bra{v} + \ket{v} \bra{u} )
\nonumber \\
&&+
(n+1) 
\left\{
\frac{1}{4(2n-1)}+
\frac
{1-p}
{ \binom{2n}{n} }
\left[1-\cfrac{1}{n}-\left(1-\cfrac{2}{n}\right)\lambda_0
\right]
\right\}
\left(\mathcal{Z}^{n-1} +\mathcal{Z}^{n+1}\right)
+
\frac{ 1 }{2n(2n-1)}
\sum_{ \substack{
u \in B_{2n,n-1} \\
v \in B_{2n,n+1} \\
u \oplus v \in B_{2n,2} 
}}
( \ket{u} \bra{v} + \ket{v} \bra{u} )
\nonumber\\
&&+
\frac
{1-p}
{ \binom{2n}{n} }
\sum_{l=1}^{n-2} 
\binom{2n-l}{n}
\left[
\cfrac{l}{n}+\left(1-\cfrac{2l}{n}\right)\lambda_0
\right]\left(\mathcal{Z}^{l} +\mathcal{Z}^{2n-l}\right).
\label{ourstrategydecomposefinalproof}
\end{eqnarray}
Since
\begin{eqnarray}
\left(
\sum_{ \substack{
u \in B_{2n,0} \\
v \in B_{2n,n} 
}}
+
\sum_{ \substack{
u \in B_{2n,2n} \\
v \in B_{2n,n} 
}}
\right)
( \ket{u} \bra{v} + \ket{v} \bra{u} )=\sqrt{2\binom{2n}{n}}\left(\ket{{\rm GHZ}_{2n}}\bra{D_{2n}^n}+\ket{D_{2n}^n}\bra{{\rm GHZ}_{2n}}\right),
\end{eqnarray}
Eq.~(\ref{ourstrategydecomposefinalproof}) shows Eq.~(\ref{M1M2M3}).
\section*{Appendix B: Eigenvalues of $\Omega^{(1)}$}
\label{sec:eigenvalue_of_Omega1}
Before deriving eigenvalues and eigenvectors of $\Omega^{(1)}$, we review the Johnson graph.
The adjacency matrix of the Johnson graph is defined as follows~\cite{BCN89}:
\begin{eqnarray}
J(n,k)
\equiv
\sum_{ \substack{
u,v \in B_{n,k} \\
u \oplus v\in B_{n,2} 
}}
\ket{u} \bra{v}.
\end{eqnarray}
The $(l+1)$th largest eigenvalue of $J(n,k)$ is
\begin{eqnarray}
\label{eq:eigenvalue-of-Johnson_graph}
\Lambda(J,n,k,l)
\equiv
k(n-k) -l(n+1-l),
\end{eqnarray}
where 
$
0\le l\le\min{ \{ k,n-k  \}  }
$.
Note that there exist $(\binom{n}{l}-\binom{n}{l-1})$ eigenvectors whose eigenvalue is $\Lambda(J,n,k,l)$.
We define an eigenvector corresponding to the eigenvalue $\Lambda(J,n,k,l)$ by
\begin{eqnarray}
\label{eq:eigenvector-of-Johnson_graph}
\ket{ \lambda (J,n,k,l,m(l))}
\equiv
\sum_{ u \in B_{n,k} } 
c_{l,m(l),u}
\ket{u},
\end{eqnarray}
where
$
1\le m(l)\le\binom{n}{l} - \binom{n}{l-1} 
$ represents the degeneracy.
Particularly when $l=0$, we can easily check
\begin{eqnarray}
\label{eq:eigenvector-of-Johnson_graph-with-largest-eigenvalue}
\ket{ \lambda (J,n,k,0,1)}
=|D_n^k\rangle
\end{eqnarray}
because $\lambda(J,n,k,0)=k(n-k)$, and the eigenvector is not degenerated.
From the fact that different eigenvectors are orthogonal,
\begin{eqnarray}
\label{eq:orthogonality-between-1st_eigenvector-and-other_eigenvector}
\sum_{ u \in B_{n,k} } 
c_{l,m(l),u}
=\sqrt{\binom{n}{k}}\braket
{
\lambda (J,n,k,0,1)
|
\lambda (J,n,k,l,m(l))
}
=
0
\end{eqnarray}
holds for any positive $l$ and $m(l)$.
\par
To obtain eigenvalues and eigenvectors of $\Omega^{(1)}$, we show the following theorem:
\begin{theorem}
\label{theorem1}
Let $n$ be any even number and $a$, $b$, $c$, and $d$ be any positive numbers.
The eigenvalues and eigenvectors $(\lambda,|\lambda\rangle)$ of the $n$-qubit Hermitian operator
\begin{eqnarray}
\hat{M}
\equiv
a \left(\mathcal{Z}^{0} +\mathcal{Z}^{n}\right)
+
b \mathcal{Z}^{\frac{n}{2}} 
+
c
\sum_{ \substack{
u,v \in B_{n,\frac{n}{2}} \\
u \oplus v\in B_{n,2} 
}}
\ket{u} \bra{v} 
+
d
\left(
\sum_{ \substack{
u \in B_{n,0} \\
v \in B_{n,\frac{n}{2}} 
}}
+
\sum_{ \substack{
u \in B_{n,n} \\
v \in B_{n,\frac{n}{2}}
}}
\right)
( \ket{u} \bra{v} + \ket{v} \bra{u} )
\end{eqnarray}
are as follows:
\begin{enumerate}
\item For $1\le l\le n/2$ and $1\le m(l)\le \binom{n}{l}-\binom{n}{l-1}$, $(\lambda,|\lambda\rangle)=(b+c\Lambda(J,n,\frac{n}{2},l),|\lambda(J,n,\frac{n}{2},l,m(l))\rangle)$,
\item $(\lambda,|\lambda\rangle)=(a,(\ket{0^n}-\ket{1^n})/\sqrt{2})$,
\item
\begin{eqnarray}
(\lambda,|\lambda\rangle)=\left(\Lambda_{\pm},\frac{\sqrt{\binom{n}{\frac{n}{2}}}\ket{D_n^{\frac{n}{2}}}+\sqrt{2}\alpha_{\pm}\ket{{\rm GHZ}_n}}{\sqrt{\binom{n}{\frac{n}{2}}+2{\alpha_{\pm}}^2}}\right)
\end{eqnarray}
with
\begin{eqnarray}
\alpha_{\pm}
\equiv
\frac
{ 4a - 4b - cn^{2} 
\pm
\sqrt{ ( 4b + cn^{2} - 4a )^{2}
+
128 d^{2} \binom{n}{\frac{n}{2}} }
}
{16d },
\end{eqnarray}
\begin{eqnarray}
\Lambda_{\pm}
\equiv
b + c \left( \frac{n}{2} \right)^{2}
+
2\alpha_{\pm} d.
\end{eqnarray}
\end{enumerate}
\end{theorem}
(Proof)
We transform $\hat{M}$ as follows:
\begin{eqnarray}
\hat{M}=
a \left(\mathcal{Z}^{0} + \mathcal{Z}^{n}\right)
+
b \mathcal{Z}^{\frac{n}{2}} 
+
c J\left(n, \frac{n}{2} \right) 
+
d
\sum_{ v \in B_{n,\frac{n}{2}} }
\left[ ( \ket{0^n} + \ket{1^n} ) \bra{v} 
+
\ket{v} ( \bra{0^n} + \bra{1^n} ) \right].
\end{eqnarray}
Therefore, from Eq.~\eqref{eq:orthogonality-between-1st_eigenvector-and-other_eigenvector},
\begin{eqnarray}
&{}&
\hat{M} 
\ket{ \lambda \left(J,n,\frac{n}{2},l,m(l)\right) }
\nonumber\\
&=&
\left\{
a \left(\mathcal{Z}^{0} + \mathcal{Z}^{n}\right)
+
b \mathcal{Z}^{\frac{n}{2}} 
+
c J\left(n, \frac{n}{2} \right) 
+
d
\sum_{ v \in B_{n,\frac{n}{2}} }
[ ( \ket{0^n} + \ket{1^n} ) \bra{v} 
+
\ket{v} ( \bra{0^n} + \bra{1^n} )]
\right\}
\left(
\sum_{ u \in B_{n,\frac{n}{2}} } 
c_{l,m(l),u} \ket{u} 
\right)\ \ \ \ \\
&=&
b 
\left(\sum_{ u \in B_{n,\frac{n}{2}} } 
c_{l,m(l),u} \ket{u}\right)
+
c \Lambda\left(J,n,\frac{n}{2},l\right)
\left(\sum_{ u \in B_{n,\frac{n}{2}} } 
c_{l,m(l),u} \ket{u}\right)
+
d\left(
\sum_{ u \in B_{n,\frac{n}{2}} }
c_{l,m(l),u} 
\right)( \ket{0^n} + \ket{1^n} )\\
&=&
\left( b + c \Lambda\left(J,n,\frac{n}{2},l\right)\right)
\ket{ \lambda \left(J,n,\frac{n}{2},l,m(l)\right) } 
\end{eqnarray}
holds for $1\le l\le n/2$ and $1\le m(l)\le\binom{n}{l}-\binom{n}{l-1}$, which shows the first case.
\par
For the second case, we can easily check
\begin{eqnarray}
\hat{M}\cfrac{\ket{0^{n}} - \ket{1^{n}}}{\sqrt{2}}
=
a\cfrac{\ket{0^{n}} - \ket{1^{n}}}{\sqrt{2}}.
\end{eqnarray}
\par
At the end of this proof, we show the third case.
From
\begin{eqnarray}
&{}&
\hat{M} 
\left[\left(\sum_{ u \in B_{n,\frac{n}{2}} }
\ket{u}\right)
+
\alpha ( \ket{0^{n}} + \ket{1^{n}} )\right]
\nonumber\\
&=&
\left\{
a \left(\mathcal{Z}^{0} + \mathcal{Z}^{n}\right)
+
b \mathcal{Z}^{\frac{n}{2}} 
+
c J\left(n, \frac{n}{2}\right) 
+
d
\sum_{ v \in B_{n,\frac{n}{2}} }
\left[ ( \ket{0^n} + \ket{1^n} ) \bra{v} 
+
\ket{v} ( \bra{0^n} + \bra{1^n} ) \right]
\right\}
\left[\left(\sum_{ u \in B_{n,\frac{n}{2}} } 
\ket{u}\right)
+
\alpha ( \ket{0^{n}} + \ket{1^{n}} )\right]\nonumber\\
\\
&=&
a \alpha ( \ket{0^{n}} + \ket{1^{n}} )
+
\left[
b + c \left( \frac{n}{2} \right)^{2}
\right]
\left(\sum_{ u \in B_{n,\frac{n}{2}} } 
\ket{u}\right) 
+
d 
\left[\sum_{ v \in B_{n,\frac{n}{2}} }
( \ket{0^n} + \ket{1^n} )\right]
+
2\alpha d
\left(\sum_{ v \in B_{n,\frac{n}{2}} }
\ket{v}\right)\\
&=&
\left[
b + c \left( \frac{n}{2} \right)^{2}
+
2\alpha d 
\right]
\left(\sum_{ u \in B_{n,\frac{n}{2}} }
\ket{u}\right) 
+
\left(
\alpha a + d \binom{n}{\frac{n}{2}}
\right)
( \ket{0^{n}} + \ket{1^{n}} ),
\label{thirdcaseeigenstateeigenvalue1}
\end{eqnarray}
we notice that a necessary condition on which $(\sum_u\in B_{n,\frac{n}{2}}\ket{u})+\alpha(|0^n\rangle+\ket{1^n})$ becomes an eigenvector is
\begin{eqnarray}
&&\alpha
\left[
b + c \left( \frac{n}{2} \right)^{2}
+
2\alpha d 
\right]
=
\alpha a + d \binom{n}{\frac{n}{2}}\\
&\iff&
8d \alpha^{2}
+
\left( 4b + cn^{2} - 4a \right)\alpha
- 
4d \binom{n}{\frac{n}{2}}
=
0.
\end{eqnarray}
The solutions of this quadratic equation are $\alpha=\alpha_{\pm}$, which mean that
\begin{eqnarray}
\cfrac{\sqrt{\binom{n}{\frac{n}{2}}}\ket{D_n^{\frac{n}{2}}}+\sqrt{2}\alpha_\pm\ket{{\rm GHZ}_n}}{\sqrt{\binom{n}{\frac{n}{2}}+2{\alpha_\pm}^2}}
\end{eqnarray}
are eigenvectors.
From Eq.~(\ref{thirdcaseeigenstateeigenvalue1}), the corresponding eigenvalues are $\Lambda_\pm$.
It is worth mentioning that all derived eigenvectors are properly orthogonal.
\hspace{\fill}$\blacksquare$
\par
When the number of qubits is $2n$, by substituting
\begin{eqnarray}
\label{abcd}
\begin{cases}
a=p + (1-p) \lambda_{0} 
\\
b =
\frac{3n-2}{2(2n-1)}
-
\frac
{ 2(1-p) \lambda_{0} }
{ \binom{2n}{n} }
\\
c = 
\frac{1}{ 2\binom{2n}{2} }
\\
d =
\frac
{ (1-p) \sqrt{ \lambda_{0} \lambda_{1} } }
{ \binom{2n}{n} },
\end{cases}
\end{eqnarray}
$\hat{M}$ in Theorem~\ref{theorem1} becomes $\Omega^{(1)}$.
Note that $a$, $b$, $c$, and $d$ in Eq.~(\ref{abcd}) are positive because $ 0 < \lambda_{1} \le 1/2 \le \lambda_{0} < 1 $ and $p<1$.
In this case, from
\begin{eqnarray}
4a - 4b - 4cn^{2}=
4 
[ p + (1-p) \lambda_{0} ]
-
4
\left[
\frac{3n-2}{2(2n-1)}
-
\frac
{ 2(1-p) \lambda_{0} }
{ \binom{2n}{n} }
\right]
-
\frac{ 4n^{2} }{ 2 \binom{2n}{2} }
=
4(1-p)
\left( \frac
{ 2 \lambda_{0} }
{ \binom{2n}{n} }
-
\lambda_{1}
\right)
\end{eqnarray}
and
\begin{eqnarray}
&{}&
\left( 4b + 4cn^{2} - 4a \right)^{2}
+
128 d^{2} \binom{2n}{n}
\nonumber \\
&=&
\left[
4(1-p)
\left( \frac
{ 2 \lambda_{0} }
{ \binom{2n}{n} }
-
\lambda_{1}
\right)
\right]
^{2}
+
128 \binom{2n}{n}
\left[
\frac
{ (1-p) \sqrt{ \lambda_{0} \lambda_{1} } }
{ \binom{2n}{n} }
\right]
^{2}
=
16(1-p)^{2}
\left(
\frac
{ 2 \lambda_{0} }
{ \binom{2n}{n} }
+
\lambda_{1}
\right)^{2},
\end{eqnarray}
we obtain
\begin{eqnarray}
\alpha_{\pm}
&=&
\frac
{ 
4(1-p)
\left( \frac
{ 2 \lambda_{0} }
{ \binom{2n}{n}  }
-
\lambda_{1}
\right)
\pm
\sqrt{ 
16(1-p)^{2}
\left(
\frac
{ 2 \lambda_{0} }
{ \binom{2n}{n}  }
+
\lambda_{1}
\right)^{2}
}
}
{16
\frac
{ (1-p) \sqrt{ \lambda_{0} \lambda_{1} } }
{ \binom{2n}{n}  }
}
=
\frac
{ 2 \lambda_{0} (1 \pm 1)
+
\lambda_{1} \binom{2n}{n}
( -1 \pm 1) }
{ 4 \sqrt{ \lambda_{0} \lambda_{1} } }.
\end{eqnarray}
The eigenvector corresponding to $\alpha_+=\sqrt{\lambda_0/\lambda_1}=\sqrt{\binom{2n}{n}q_0/(2q_1)}$ in the third case is our target state Eq.~\eqref{superposition}, and hence its eigenvalue is
\begin{eqnarray}
\Lambda\left( \Omega^{(1)}, \Lambda_{+} \right)\equiv\Lambda_+
=
b + cn^{2}
+
2\alpha_{+} d 
=\cfrac{3n-2}{2(2n-1)}-\cfrac{2(1-p)\lambda_0}{\binom{2n}{n}}+\cfrac{n^2}{2n(2n-1)}+2\cfrac{(1-p)\sqrt{\lambda_0\lambda_1}}{\binom{2n}{n}}\sqrt{\cfrac{\lambda_0}{\lambda_1}}
=
1.
\end{eqnarray}
On the other hand, the eigenvector corresponding to $\alpha_-=-\sqrt{\binom{2n}{n}q_1/(2q_0)}$ is $\sqrt{q_1}\ket{{\rm GHZ}_{2n}}-\sqrt{q_0}\ket{D_{}2n^n}$, and its eigenvalue $\Lambda(\Omega^{(1)},\Lambda_-)\equiv\Lambda_-$ is
\begin{eqnarray}
\Lambda\left( \Omega^{(1)}, \Lambda_- \right)
&=&
b + cn^2
+
2\alpha_- d\\
&=&
\frac
{ 4a + 4b + 4cn^{2} 
-
\sqrt{ ( 4b + 4cn^{2} - 4a )^{2}
+
128 d^{2} \binom{2n}{n} }
}
{8 }\\
&=&
\frac
{ 
4(1+p)
+
4(1-p) \lambda_{0}
\left( 
1 - \frac{2}{ \binom{2n}{n} }
\right)
-
\sqrt{ 
16(1-p)^{2}
\left(
\frac
{ 2 \lambda_{0} }
{ \binom{2n}{n} }
+
\lambda_{1}
\right)^{2}
}
}
{8 }\\
&=&
\frac{1}{2}
\left\{
1+p
+
(1-p) 
\left[
2\lambda_{0}
\left( 
1 - \frac{2}{ \binom{2n}{n} }
\right)
-
1 
\right]
\right\}\\
&=&
p
+
\lambda_{0} (1-p) 
\left( 
1 - \frac{2}{ \binom{2n}{n} }
\right).
\end{eqnarray}
\par
The eigenvalue in the second case is
\begin{eqnarray}
\Lambda\left( \Omega^{(1)},a\right)
\equiv
a
=
p + (1-p) \lambda_{0}.
\end{eqnarray}
\par
The eigenvalues in the first case are
\begin{eqnarray}
\Lambda\left( \Omega^{(1)},b,c,l\right)
&\equiv&
b + c \Lambda(J,2n, n ,l)\\
&=&
\frac{ 3n-2 }{ 2(2n-1) }
-
\frac
{ 2 (1-p) \lambda_{0} }
{ \binom{2n}{n} }
+
\frac{1}{ 2\binom{2n}{2} }
\left[n^2 -l(2n+1-l) 
\right]\\
&=&
1
-
\frac
{ 2 (1-p) \lambda_{0} }
{ \binom{2n}{n} }
-
\frac{ l(2n+1-l) }{ 2n(2n-1) }.
\end{eqnarray}
\par
Since
\begin{eqnarray}
{\rm max}_{1\le l\le n}\Lambda\left(\Omega^{(1)},b,c,l\right)=\Lambda\left(\Omega^{(1)},b,c,1\right)=1
-
\frac
{ 2 (1-p) \lambda_{0} }
{ \binom{2n}{n} }
-
\frac{1}{2n-1},
\end{eqnarray}
the candidates of the second largest eigenvalue of 
$ \Omega^{(1)}$
are 
$
\Lambda(\Omega^{(1)},\Lambda_{-})$, $\Lambda( \Omega^{(1)},a) 
$,
and
$
\Lambda( \Omega^{(1)},b,c,1)
$.
To identify which one is the second largest eigenvalue, we first compare $ \Lambda( \Omega^{(1)}, \Lambda_{-} ) $
and
$ \Lambda( \Omega^{(1)},a) $ as follows:
\begin{eqnarray}
&&\Lambda\left( \Omega^{(1)}, \Lambda_{-} \right)
-
\Lambda\left( \Omega^{(1)},a\right)
=
- \lambda_{0} (1-p) 
\frac{2}{ \binom{2n}{n} }
<
0\\
\label{comparisonlambdaa}
&\iff&\Lambda\left( \Omega^{(1)}, \Lambda_{-} \right)
<
\Lambda\left( \Omega^{(1)},a\right).
\end{eqnarray}
Then we compare $ \Lambda( \Omega^{(1)},a) $ 
and
$ \Lambda( \Omega^{(1)},b,c,1) $.
From
\begin{eqnarray}
\Lambda\left( \Omega^{(1)},b,c,1\right)
-
\Lambda\left( \Omega^{(1)},a\right)
=
(1-p) 
\left\{
1
-
\left[
\lambda_{0} 
\left(
1 + \frac{2}{ \binom{2n}{n} }
\right)
+ 
\frac{1}{(2n-1)(1-p)}
\right]
\right\},
\end{eqnarray}
their relation depends on the value of $\lambda_0$ as follows:
\begin{eqnarray}
\label{comparisonabc1}
\begin{cases}
\Lambda\left( \Omega^{(1)},a\right) \ge \Lambda\left( \Omega^{(1)},b,c,1\right)
\quad \left( 
1 
\le
\lambda_{0} 
\left(
1 + \frac{2}{ \binom{2n}{n} }
\right)
+ 
\frac{1}{(2n-1)(1-p)}
\right)
\\
\Lambda\left( \Omega^{(1)},a\right) < \Lambda\left( \Omega^{(1)},b,c,1\right)
\quad \left({\rm otherwise}\right).
\end{cases}
\end{eqnarray}
By combining Eq.~(\ref{comparisonabc1}) with Eq.~(\ref{comparisonlambdaa}), we finally derive
\begin{eqnarray}
\beta ( \Omega^{(1)})
=
\max
\left\{
\Lambda\left( \Omega^{(1)},a\right),\Lambda\left( \Omega^{(1)},b,c,1\right)
\right\}=
\begin{cases}
\Lambda\left( \Omega^{(1)},a\right) 
\quad \left( 
1 
\le
\lambda_{0} 
\left(
1 + \frac{2}{ \binom{2n}{n} }
\right)
+ 
\frac{1}{(2n-1)(1-p)}
\right)
\\
\Lambda\left( \Omega^{(1)},b,c,1\right)
\quad \left({\rm otherwise}\right).
\end{cases}
\end{eqnarray}
\section*{Appendix C: The largest eigenvalue of $\Omega^{(2)}$}
\label{sec:largest_eigenvalue_of_Omega2}
To obtain the largest eigenvalue of $\Omega^{(2)}$, we show the following theorem:
\begin{theorem}
\label{theorem2}
Let $\alpha$ and $\beta$ be any positive numbers.
For the $n$-qubit nonnegative and irreducible operator
\begin{eqnarray}
\label{eq:operator-gives-Omega2}
\hat{M}
\equiv
\alpha \left(\mathcal{Z}^{k-1} + \mathcal{Z}^{k+1} \right)
+
\beta
\sum_{ \substack{
u \in B_{n,k-1} \\
v \in B_{n,k+1} \\
u \oplus v\in B_{n,2} 
}}
( \ket{u} \bra{v} + \ket{v} \bra{u} ),
\end{eqnarray}
the maximum eigenvalue $\Lambda_1(\hat{M})$ and the corresponding eigenvector are
\begin{eqnarray}
\label{eq:eigenvalue-of-operator-gives-Omega2}
\Lambda_{1}( \hat{M} )
=
\alpha
+
\beta \sqrt{ \binom{n-k+1}{2}\binom{k+1}{2} }
\end{eqnarray}
and
\begin{eqnarray}
\label{eq:eigenvector-of-operator-gives-Omega2}
\ket{\lambda}
\equiv
\frac{\ket{D_{n}^{k-1}} + \ket{D_{n}^{k+1}}}{\sqrt{2}},
\end{eqnarray}
respectively.
Note that $\hat{M}$ is treated as an operator on the space spanned by $\{\ket{u}\}_{u\in B_{n,k-1}\cup B_{n,k+1}}$.
\end{theorem}
(Proof)
The proof is essentially the same as that used in Ref.~\cite{LYSZZ19}.
We can easily check
\begin{eqnarray}
\hat{M}\cfrac{\ket{D_{n}^{k-1}} + \ket{D_{n}^{k+1}}}{\sqrt{2}}&=&\alpha\cfrac{\ket{D_{n}^{k-1}} + \ket{D_{n}^{k+1}}}{\sqrt{2}}+\cfrac{\beta}{\sqrt{2}}\left(\sqrt{\cfrac{\binom{n}{k+1}}{\binom{n}{k-1}}}\binom{k+1}{2}\ket{D_n^{k+1}}+\sqrt{\cfrac{\binom{n}{k-1}}{\binom{n}{k+1}}}\binom{n-k+1}{2}\ket{D_n^{k-1}}\right)\ \ \ \\
&=&\left(\alpha+\beta\sqrt{\binom{n-k+1}{2}\binom{k+1}{2}}\right)\cfrac{\ket{D_{n}^{k-1}} + \ket{D_{n}^{k+1}}}{\sqrt{2}},
\end{eqnarray}
and hence $|\lambda\rangle$ is the eigenvector with the eigenvalue $\Lambda_1(\hat{M})$.
From the Perron-Frobenius theorem (see Chapter 8 in Ref.~\cite{meyer2000matrix}), the eigenvalue $\Lambda_1(\hat{M})$ 
is the largest one, and $\ket{\lambda}$ is nondegenerate.
\hspace{\fill}$\blacksquare$
\par
By replacing the number $n$ of qubits with $2n$ and substituting
\begin{eqnarray}
\begin{cases}
k=n
\\
\alpha
=
(n+1) 
\left\{
\cfrac
{1-p}
{ \binom{2n}{n} }
\left[
\lambda_{0} 
-
\cfrac{n-1}{n}
( 2 \lambda_{0} -1 )
\right]
+
\cfrac{ n }{4\binom{2n}{2}}
\right\}
\\
\beta
=
\cfrac{1}{ 2 \binom{2n}{2} }
\end{cases}
\end{eqnarray}
into $\hat{M}$ in Theorem~\ref{theorem2}, we derive the largest eigenvalue $\Lambda_1(\Omega^{(2)})$ as
\begin{eqnarray}
\Lambda_{1}\left( \Omega^{(2)} \right)
=
\alpha
+
\beta\binom{n+1}{2}
=
\cfrac{1}{4} + \cfrac{3}{ 4(2n-1) }
+
\cfrac
{ (1-p)(n+1) }
{ \binom{2n}{n} }
\left[
\lambda_{0} 
\left(
\cfrac{2}{n} -1 
\right)
+
1 - \cfrac{1}{n}
\right].
\end{eqnarray}
\section*{Appendix D: The largest eigenvalue of $\Omega^{(3)}$}
\label{sec:largest_eigenvalue_of_Omega3}
From Eq.~\eqref{eq:Omega3}, the eigenvalues of $\Omega^{(3)}$ are $\{\Lambda(\Omega^{(3)},l)\}_{l=1}^{n-2}$, where
\begin{eqnarray}
\Lambda\left( \Omega^{(3)} ,l \right)
\equiv
\cfrac
{ 1-p }
{ \binom{2n}{n} }
\binom{2n-l}{n} 
\left[
\lambda_{0} 
-
\cfrac{l}{n}
( 2 \lambda_{0} -1 )
\right]
=
\cfrac
{ 1-p }
{ n \binom{2n}{n} }
\binom{2n-l}{n} 
[ n \lambda_{0} 
-
l ( 2 \lambda_{0} -1 ) ].
\end{eqnarray}
To derive ${\rm max}_{1\le l\le n-2}\Lambda(\Omega^{(3)},l)$, we evaluate the ratio
\begin{eqnarray}
\Lambda_{r} \left( \Omega^{(3)} ,n,l \right) 
&\equiv&
\cfrac
{ \Lambda\left( \Omega^{(3)} ,l+1 \right) }
{ \Lambda\left( \Omega^{(3)} ,l \right) }\\
&=&
\cfrac{
\binom{2n-l-1}{n} 
[ n \lambda_{0} 
-
(l+1) ( 2 \lambda_{0} -1 ) ]}
{\binom{2n-l}{n} 
[ n \lambda_{0} 
-
l ( 2 \lambda_{0} -1 ) ]}\\
&=&
\left(
1 - \cfrac{n} { 2n-l }
\right)
\left[
1
-
\cfrac
{ 2 \lambda_{0} -1 }
{ ( n-2l )\lambda_{0} + l }
\right].
\end{eqnarray}
Since
\begin{eqnarray}
\cfrac{\partial}{\partial \lambda_0}\left[\cfrac
{ 2 \lambda_{0} -1 }
{ ( n-2l )\lambda_{0} + l }\right]=\cfrac{n}{\left[(n-2l)\lambda_0+l\right]^2}>0,
\end{eqnarray}
the ratio $\Lambda_{r} ( \Omega^{(3)} ,n,l )$ takes its maximum when $\lambda_0=1/2$, and hence
\begin{eqnarray}
&&\Lambda_{r} \left( \Omega^{(3)} ,n,l \right)\le 1-\cfrac{n}{2n-l}
<
1\\
&\iff&\Lambda\left( \Omega^{(3)},l+1 \right) 
<
\Lambda\left( \Omega^{(3)},l \right).
\end{eqnarray}
Therefore,
the largest eigenvalue $\Lambda_1(\Omega^{(3)})$ of 
$\Omega^{(3)}$
is
\begin{eqnarray}
\Lambda_{1} \left( \Omega^{(3)} \right) 
=
\Lambda\left( \Omega^{(3)} ,1\right) 
=
\cfrac
{ 1-p }
{ n \binom{2n}{n} }
\binom{2n-1}{n} 
\left[ n \lambda_{0} 
-( 2 \lambda_{0} -1 ) \right]
=
\cfrac{1}{2n}
(1-p)[ \lambda_{0} ( n-2) +1 ].
\end{eqnarray}
\section*{Appendix E: Derivation of $\beta(\Omega)$ }
To derive $\beta(\Omega)$, we compare $\Lambda_1(\Omega^{(2)})$ and $\Lambda_1(\Omega^{(3)})$ with $\Lambda(\Omega^{(1)},b,c,1)$ and $\Lambda(\Omega^{(1)},a)$ one by one.
\setcounter{subsection}{0}
\subsection{
Comparison between 
$ \Lambda_{1}( \Omega^{(2)} ) $ 
and
$ \Lambda( \Omega^{(1)},b,c,1) $ 
}
We obtain
\begin{eqnarray}
\label{eq:diff-of-eigenvalue-between-Omega2-and-Omega1_bc}
&{}&
\Lambda_{1}\left( \Omega^{(2)} \right)
-
\Lambda\left( \Omega^{(1)} ,b,c,1\right)
\nonumber\\
&=&
\frac{1}{4} + \frac{3}{ 4(2n-1) }
+
\frac
{ (1-p)(n+1) }
{ \binom{2n}{n} }
\left[
\lambda_{0} 
\left(
\frac{2}{n} -1 
\right)
+
1 - \frac{1}{n}
\right]
-
\left[
1
-
\frac
{ 2 (1-p) \lambda_{0} }
{ \binom{2n}{n} }
-
\frac{1}{2n-1}
\right]\\
&=&
\frac{7}{ 4(2n-1) } - \frac{3}{4}
+
\frac
{ 1-p }
{ \binom{2n}{n} }
\lambda_{0}
\left(
\frac{2}{n} + 3 - n
\right)
+
\frac
{1-p}
{ \binom{2n}{n} }
\left(
n - \frac{1}{n}
\right).
\label{comparison21bc12}
\end{eqnarray}
Since
\begin{eqnarray}
\left(\cfrac{2}{n}+3-n\right)-\left[\cfrac{2}{n+1}+3-(n+1)\right]=1+\cfrac{2}{n(n+1)}>0,
\end{eqnarray}
Eq.~(\ref{comparison21bc12}) takes its maximum when $\lambda_0=1$ and $\lambda_0=1/2$ when $n=3$ and $n\ge 4$, respectively.
Therefore, when $n=3$,
\begin{eqnarray}
&{}&
\Lambda_{1}\left( \Omega^{(2)} \right)
-
\Lambda\left( \Omega^{(1)} ,b,c,1\right)\le
\cfrac{7}{20} - \cfrac{3}{4}
+
\cfrac
{2 (1-p) }
{ 3\binom{6}{3} }
+
\cfrac
{1-p}
{ \binom{6}{3} }
\left(
3 - \cfrac{1}{3}
\right)\le-\cfrac{7}{30}<0.
\end{eqnarray}
\par
On the other hand, when $n\ge4$,
\begin{eqnarray}
&{}&
\Lambda_{1}\left( \Omega^{(2)} \right)
-
\Lambda\left( \Omega^{(1)} ,b,c,1\right)
\nonumber\\
&
\le
&
\frac{7}{ 4(2n-1) } - \frac{3}{4}
+
\frac
{ 1-p }
{ 2\binom{2n}{n} }
\left(
\frac{2}{n} + 3 - n
\right)
+
\frac
{ 1-p }
{ \binom{2n}{n} }
\left(
n - \frac{1}{n}
\right)\\
&=&
\frac{7}{ 4(2n-1) } - \frac{3}{4}
+
\frac
{ (1-p)(n+3) }
{ 2\binom{2n}{n} }\\
&
\le
&
\frac{7}{ 4(8-1) } - \frac{3}{4}
+
\frac
{ (1-p)(4+3) }
{ 2\binom{8}{4} }\\
&=&
- \frac{ p+9 }{ 20 } < 0,
\end{eqnarray}
where we have used the fact that $(n+3)/\binom{2n}{n}$ is a monotonically decreasing function in $n$ to derive the second inequality.
\par
In short, we conclude
\begin{eqnarray}
\label{comparison21bc1}
\Lambda_{1}\left( \Omega^{(2)}\right)
<
\Lambda\left( \Omega^{(1)} ,b,c,1\right).
\end{eqnarray}
\subsection{
Comparison between 
$ \Lambda_{1}( \Omega^{(2)} ) $ 
and
$\Lambda( \Omega^{(1)},a)$
}
We show
\begin{eqnarray}
&{}&
\Lambda_{1}\left( \Omega^{(2)} \right)
-
\Lambda\left( \Omega^{(1)},a\right)
\nonumber\\
&=&
\frac{1}{4} + \frac{3}{ 4(2n-1) }
+
\frac
{ (1-p)(n+1) }
{ \binom{2n}{n} }
\left[
\lambda_{0} 
\left(
\frac{2}{n} -1 
\right)
+
1 - \frac{1}{n}
\right]
-
[ p + (1-p) \lambda_{0} ]\\
&
\le
&
\frac{1}{4} + \frac{3}{ 4(2n-1) }
+
\frac
{ (1-p)(n+1) }
{ \binom{2n}{n} }
\left[
\frac{1}{2}
\left(
\frac{2}{n} -1 
\right)
+
1 - \frac{1}{n}
\right]
-
\left[
p + (1-p) \frac{1}{2} 
\right]\\
&=&
\frac{1}{4} + \frac{3}{ 4(2n-1) }
+
\frac
{ (1-p)(n+1) }
{ 2\binom{2n}{n} }
-
\frac{1+p}{2}\\
&
\le
&
\frac{1}{4} + \frac{3}{ 4(2n-1) }
+
\frac
{ n+1 }
{ 2\binom{2n}{n} }
-
\frac{1}{2} 
\le
\frac{3}{20}
+
\frac
{4}
{ 2\binom{6}{3} }
-
\frac{1}{4} 
=
0,
\end{eqnarray}
where we have used $\lambda_0\ge1/2$, $p\ge0$, and $n\ge3$ to derive the first, second, and third inequalities.
In conclusion,
\begin{eqnarray}
\label{comparison21a}
\Lambda_{1}\left( \Omega^{(2)} \right)
\le
\Lambda\left( \Omega^{(1)},a\right).
\end{eqnarray}
%ここから
\subsection{Comparison between 
$ \Lambda_{1}( \Omega^{(3)} ) $ 
and
$ \Lambda( \Omega^{(1)} ,b,c,1) $ 
}
We obtain
\begin{eqnarray}
&{}&
\Lambda\left( \Omega^{(1)},b,c,1\right)
-
\Lambda_{1} \left( \Omega^{(3)} \right) 
\nonumber\\
&=&
1
-
\frac
{ 2 (1-p) \lambda_{0} }
{ \binom{2n}{n} }
-
\frac{1}{2n-1}
-
\frac{1}{2n}
(1-p)[ \lambda_{0} ( n-2) +1 ]\\
&=&
1 
- 
\left[
\frac{1}{2n-1} + \frac{1-p}{2n}
+
(1-p) \lambda_{0}
\left(
\frac{n-2}{2n}
+
\frac{2}{ \binom{2n}{n} }
\right)
\right]\\
&
>
&
1 
- 
\left[
\frac{1}{2n-1} + \frac{1-p}{2n}
+
(1-p) 
\left(
\frac{n-2}{2n}
+
\frac{2}{ \binom{2n}{n} }
\right)
\right]\\
&
\ge
&
1 
- 
\left(
\frac{1}{2n-1} + \frac{1}{2n}
+
\frac{n-2}{2n}
+
\frac{2}{ \binom{2n}{n} }
\right),
\end{eqnarray}
where we have used $\lambda_0<1$, $n\ge 3$, and $p\ge0$ to derive the first and second inequalities.
Since
\begin{eqnarray}
\frac{1}{2n-1} + \frac{1}{2n}
+
\frac{n-2}{2n}
+
\frac{2}{ \binom{2n}{n} }
=
\frac{1}{2n(2n-1)} + \frac{1}{2}
+
\frac{2}{ \binom{2n}{n} }
\le
\frac{1}{30} + \frac{1}{2}
+
\frac{2}{ \binom{6}{3} }
=
\frac{19}{30}
<
1,
\end{eqnarray}
where we have used that $\binom{2n}{n}$ is monotonically increasing in $n$ and $n\ge3$ to derive the first inequality, we conclude
\begin{eqnarray}
\label{comparison31bc1}
\Lambda_{1} \left( \Omega^{(3)} \right) 
<
\Lambda\left( \Omega^{(1)},b,c,1\right).
\end{eqnarray}
\subsection{
Comparison between 
$ \Lambda_{1}( \Omega^{(3)} ) $ 
and
$ \Lambda( \Omega^{(1)},a) $
}
From
\begin{eqnarray}
\Lambda_{1}\left( \Omega^{(3)} \right)
-
\Lambda\left( \Omega^{(1)},a\right)
&=&
\frac{1}{2n}
[ 1-p(2n+1) - (1-p) \lambda_{0} (n+2) ]\\
&
\le
&
\frac{1}{2n}
\left[ 1-p(2n+1) - \frac{1}{2} (1-p) (n+2)\right]\\
&=&
-\frac{3p+1}{4}
<
0,
\end{eqnarray}
where we have used $\lambda_0\ge1/2$ to derive the first inequality, we obtain
\begin{eqnarray}
\label{comaprison31a}
\Lambda_{1} \left( \Omega^{(3)} \right)
<
\Lambda\left( \Omega^{(1)},a\right).
\end{eqnarray}
\subsection{Summary}
From Eqs.~(\ref{comparison21bc1}), (\ref{comparison21a}), (\ref{comparison31bc1}), and (\ref{comaprison31a}), the second largest eigenvalue $\beta(\Omega)$ of our strategy $\Omega$ is
\begin{eqnarray}
\beta \left(\Omega\right)= 
\beta \left(\Omega^{(1)}\right)=
\begin{cases}
\Lambda\left( \Omega^{(1)},a\right)\quad 
\left(
1 
\le
\lambda_{0} 
\left(
1 + \frac{2}{ \binom{2n}{n} }
\right)
+ 
\frac{1}{(2n-1)(1-p)}
\right)
\\
\Lambda\left( \Omega^{(1)},b,c,1\right)
\quad \left({\rm otherwise}\right).
\end{cases}
\end{eqnarray}
\end{document}